\documentclass[twocolumn]{aastex631}

\usepackage{savesym}
\savesymbol{tablenum}
\usepackage{siunitx}
\restoresymbol{SIX}{tablenum}

\usepackage[version=4]{mhchem}
\usepackage{graphicx,epsf}
\usepackage{subfigure}
\usepackage{graphicx}	
\usepackage{amsmath}	
\usepackage{amssymb}	
\usepackage{newtxtext,newtxmath}
\usepackage{siunitx}
\usepackage{footnote}

\newcommand{\Msun}{{\rm M}_\odot}
\newcommand{\Rsun}{{\rm R}_\odot}

\newcommand{\kms}{\textrm{km}\,\textrm{s}^{-1}}

\def\sn{{SN~2023ixf}}

\newcommand{\NSF}{NSF Graduate Research Fellow}

\DeclareRobustCommand{\ion}[2]{\relax\ifmmode\ifx\testbx\f@series{\mathbf{#1\,\mathsc{#2}}}\else{\mathrm{#1\,\mathsc{#2}}}\fi\else\textup{#1\,{\mdseries\textsc{#2}}}\fi}

\newcommand{\code}[1]{\texttt{#1}}
\def\heracles{{\code{HERACLES}}}
\def\cmfgen{{\code{CMFGEN}}}

\usepackage[scale=0.94]{tgbonum}
\usepackage[mode=text]{siunitx}

\makeatletter
\DeclareTextCompositeCommand{\r}{OT1}{A}{%
  \leavevmode\vbox{%
    \offinterlineskip
    \ialign{\hfil##\hfil\cr\char23\cr\noalign{\kern-1.15ex}A\cr}%
  }%
}
\makeatother

\shorttitle{Supernova 2023ixf}
\shortauthors{Jacobson-Gal\'an et al.}

\begin{document}

\title{SN~2023ixf in Messier 101: Photo-ionization of Dense, Close-in Circumstellar Material in a Nearby Type II Supernova}


\correspondingauthor{Wynn Jacobson-Gal\'{a}n (he, him, his)}
\email{wynnjg@berkeley.edu}

\author[0000-0002-3934-2644]{W.~V.~Jacobson-Gal\'{a}n}
\affil{Department of Astronomy, University of California, Berkeley, CA 94720, USA}
\affil{\NSF}

\author[0000-0003-0599-8407]{L.~Dessart}
\affil{Institut d’Astrophysique de Paris, CNRS-Sorbonne Université, 98 bis boulevard Arago, F-75014 Paris, France}

\author[0000-0003-4768-7586]{R.~Margutti}
\affil{Department of Astronomy, University of California, Berkeley, CA 94720, USA}
\affil{Department of Physics, University of California, Berkeley, CA 94720, USA}

\author[0000-0002-7706-5668]{R.~Chornock}
\affil{Department of Astronomy, University of California, Berkeley, CA 94720, USA}

\author[0000-0002-2445-5275]{R.~J.~Foley}
\affiliation{Department of Astronomy and Astrophysics, University of California, Santa Cruz, CA 95064, USA}

\author[0000-0002-5740-7747]{C.~D.~Kilpatrick}
\affil{Center for Interdisciplinary Exploration and Research in Astrophysics (CIERA), and Department of Physics and Astronomy, Northwestern University, Evanston, IL 60208, USA}

\author[0000-0002-6230-0151]{D.~O.~Jones}
\affiliation{Gemini Observatory, NSF’s NOIRLab, 670 N. A’ohoku Place, Hilo, HI 96720, USA}

\author[0000-0002-5748-4558]{K.~Taggart}
\affil{Department of Astronomy and Astrophysics, University of California, Santa Cruz, CA 95064, USA}

\author[0000-0002-4269-7999]{C.~R.~Angus}
\affiliation{DARK, Niels Bohr Institute, University of Copenhagen, Jagtvej 128, 2200 Copenhagen, Denmark}

\author[0000-0002-7350-7043]{S.~Bhattacharjee}
\affil{Graduate Institute of Astronomy, National Central University, 300 Zhongda Road, Zhongli, Taoyuan 32001, Taiwan}

\author[0009-0000-2983-1017]{L.~A.~Braff}
\affil{Department of Astronomy and Astrophysics, University of California, Santa Cruz, CA 95064, USA}

\author[0000-0001-6415-0903]{D.~Brethauer}
\affil{Department of Astronomy, University of California, Berkeley, CA 94720, USA}

\author[0000-0002-6523-9536]{A.~J.~Burgasser}
\affiliation{Department of Astronomy \& Astrophysics, UC San Diego, La Jolla, CA 92093, USA}

\author{F.~Cao}
\affil{Department of Statistics, University of California, Riverside, CA 92521, USA}

\author{C.~M.~Carlile}
\affil{Center for Astrophysics and Space Sciences, University of California, San Diego, La Jolla, CA 92093, USA}

\author[0000-0001-6965-7789]{K.~C.~Chambers}
\affil{Institute for Astronomy, University of Hawaii, 2680 Woodlawn Drive, Honolulu, HI 96822, USA}

\author[0000-0003-4263-2228]{D.~A.~Coulter}
\affil{\NSF}
\affil{Department of Astronomy and Astrophysics, University of California, Santa Cruz, CA 95064,
USA}

\author[0009-0009-5524-8525]{E.~Dominguez-Ruiz}
\affil{Department of Astronomy and Astrophysics, University of California, Santa Cruz, CA 95064, USA}

\author[0000-0001-9749-4200]{C.~B.~Dickinson}
\affiliation{Department of Astronomy and Astrophysics, University of California, Santa Cruz, CA 95064, USA}

\author[0000-0001-5486-2747]{T.~de~Boer}
\affil{Institute for Astronomy, University of Hawaii, 2680 Woodlawn Drive, Honolulu, HI 96822, USA}

\author[0000-0003-4906-8447]{A.~Gagliano}
\affil{\NSF}
\affil{Department of Astronomy, University of Illinois at Urbana-Champaign, 1002 W. Green St., IL 61801, USA}
\affil{Center for Astrophysical Surveys, National Center for Supercomputing Applications, Urbana, IL, 61801, USA}

\author[0000-0002-8526-3963]{C.~Gall}
\affil{DARK, Niels Bohr Institute, University of Copenhagen, Jagtvej 128, 2200 Copenhagen, Denmark}

\author[0000-0003-1015-5367]{H.~Gao}
\affil{Institute for Astronomy, University of Hawaii, 2680 Woodlawn Drive, Honolulu, HI 96822, USA}

\author[0000-0002-3739-0423]{E.~L.~Gates}
\affil{University of California Observatories/Lick Observatory, Mount Hamilton, CA 95140}

\author[0000-0001-6395-6702]{S.~Gomez}
\affil{Space Telescope Science Institute, Baltimore, MD 21218}

\author[0000-0002-0786-7307]{M.~Guolo}
\affiliation{Department of Physics and Astronomy, The Johns Hopkins University, Baltimore, MD 21218, USA}

\author[0000-0002-5440-2350]{M.~R.~J.~Halford}
\affiliation{Department of Physics, Auburn University, 380 Duncan Drive, Auburn, AL 36849, USA}

\author[0000-0002-4571-2306]{J.~Hjorth}
\affiliation{DARK, Niels Bohr Institute, University of Copenhagen, Jagtvej 128, 2200 Copenhagen, Denmark}

\author[0000-0003-1059-9603]{M.~E.~Huber}
\affiliation{Institute for Astronomy, University of Hawaii, 2680 Woodlawn Drive, Honolulu, HI 96822, USA}

\author[0009-0008-2688-0815]{M.~N.~Johnson}
\affiliation{Department of Physics, University of California, Santa Cruz, CA 95064, USA}

\author{P.~R.~Karpoor}
\affil{Center for Astrophysics and Space Sciences, University of California, San Diego, La Jolla, CA 92093, USA}

\author[0000-0003-1792-2338]{T.~Laskar}
\affil{Department of Physics \& Astronomy, University of Utah, Salt Lake City, UT 84112, USA}

\author[0000-0002-2249-0595]{N~LeBaron}
\affil{Department of Astronomy, University of California, Berkeley, CA 94720, USA}

\author[0000-0002-4860-7667]{Z.~Li}
\affiliation{Department of Earth and Planetary Sciences, University of California, Riverside, CA 92521, USA}

\author{Y.~Lin}
\affil{Department of Physics \& Astronomy, University of California Riverside, 900 University Ave, Riverside, CA}

\author[0000-0002-3822-6756]{S.~D.~Loch}
\affiliation{Department of Physics, Auburn University, 380 Duncan Drive, Auburn, AL 36849, USA}

\author{P.~D.~Lynam}
\affil{University of California Observatories/Lick Observatory, Mount Hamilton, CA 95140}

\author[0000-0002-7965-2815]{E.~A.~Magnier}
\affil{Institute for Astronomy, University of Hawaii, 2680 Woodlawn Drive, Honolulu, HI 96822, USA}

\author{P.~Maloney}
\affil{Department of Astronomy and Astrophysics, University of California, Santa Cruz, CA 95064, USA}

\author[0000-0002-4513-3849]{D.J.~Matthews}
\affil{Department of Astronomy, University of California, Berkeley, CA 94720, USA}

\author{M.~McDonald}
\affil{Department of Physics \& Astronomy, University of California Riverside, 900 University Ave, Riverside, CA}

\author[0000-0003-2736-5977]{H.-Y.~Miao}
\affil{Graduate Institute of Astronomy, National Central University, 300 Zhongda Road, Zhongli, Taoyuan 32001, Taiwan}

\author[0000-0002-0763-3885]{D. Milisavljevic}
\affil{Department of Physics and Astronomy, Purdue University, 525 Northwestern Avenue, West Lafayette, IN 47907, USA}

\author[0000-0001-8415-6720]{Y.-C.~Pan}
\affil{Graduate Institute of Astronomy, National Central University, 300 Zhongda Road, Zhongli, Taoyuan 32001, Taiwan}

\author{S.~Pradyumna}
\affil{Department of Physics \& Astronomy, University of California Riverside, 900 University Ave, Riverside, CA}

\author[0000-0003-4175-4960]{C.~L.~Ransome}
\affiliation{Department of Astronomy and Astrophysics, Pennsylvania State University, 525 Davey Laboratory, University Park, PA 16802, USA}

\author[0000-0002-5376-3883]{J.~M.~Rees}
\affil{University of California Observatories/Lick Observatory, Mount Hamilton, CA 95140}

\author[0000-0002-4410-5387]{A.~Rest}
\affil{Space Telescope Science Institute, Baltimore, MD 21218}
\affiliation{Department of Physics and Astronomy, The Johns Hopkins University, Baltimore, MD 21218, USA}

\author[0000-0002-7559-315X]{C.~Rojas-Bravo}
\affiliation{Department of Astronomy and Astrophysics, University of California, Santa Cruz, CA 95064, USA}

\author[0000-0002-7393-3595]{N.~R.~Sandford}
\affiliation{Department of Astronomy, University of California, Berkeley, CA 94720, USA}

\author[0000-0001-8568-8729]{L.~Sandoval~Ascencio}
\affiliation{Department of Physics and Astronomy, University of California, Irvine, California 92697-4575, USA}

\author{S.~Sanjaripour}
\affil{Department of Physics \& Astronomy, University of California Riverside, 900 University Ave, Riverside, CA}

\author[0000-0002-1445-4877]{A.~Savino}
\affiliation{Department of Astronomy, University of California, Berkeley, CA 94720, USA}

\author[0000-0001-8023-4912]{H.~Sears}
\affil{Center for Interdisciplinary Exploration and Research in Astrophysics (CIERA), and Department of Physics and Astronomy, Northwestern University, Evanston, IL 60208, USA}

\author{N.~Sharei}
\affil{Department of Physics \& Astronomy, University of California Riverside, 900 University Ave, Riverside, CA}

\author[0000-0002-8229-1731]{S.~J.~Smartt}
\affil{Astrophysics Research Centre, School of Mathematics and Physics, Queen’s University Belfast, Belfast BT7 1NN, UK}
\affil{Department of Physics, University of Oxford, Keble Road, Oxford, UK}

\author[0000-0002-1420-1837]{E.~R.~Softich}
\affil{Center for Astrophysics and Space Sciences, University of California, San Diego, La Jolla, CA 92093, USA}

\author[0000-0002-9807-5435]{C.~A.~Theissen}
\affil{Center for Astrophysics and Space Sciences, University of California, San Diego, La Jolla, CA 92093, USA}

\author[0000-0002-1481-4676]{S.~Tinyanont}
\affil{Department of Astronomy and Astrophysics, University of California, Santa Cruz, CA 95064, USA}

\author{H.~Tohfa}
\affil{Department of Physics \& Astronomy, University of California Riverside, 900 University Ave, Riverside, CA}

\author[0000-0002-5814-4061]{V.~A.~Villar}
\affiliation{Department of Astronomy and Astrophysics, Pennsylvania State University, 525 Davey Laboratory, University Park, PA 16802, USA}
\affiliation{Institute for Computational \& Data Sciences, The Pennsylvania State University, University Park, PA, USA}
\affiliation{Institute for Gravitation and the Cosmos, The Pennsylvania State University, University Park, PA 16802, USA}

\author[0000-0001-5233-6989]{Q.~Wang}
\affiliation{Department of Physics and Astronomy, The Johns Hopkins University, Baltimore, MD 21218, USA}

\author[0000-0002-1341-0952]{R.~J.~Wainscoat}
\affil{Institute for Astronomy, University of Hawaii, 2680 Woodlawn Drive, Honolulu, HI 96822, USA}

\author[0009-0003-8229-0127]{A.~L.~Westerling}
\affil{Department of Astronomy and Astrophysics, University of California, Santa Cruz, CA 95064, USA}

\author{E.~Wiston}
\affil{Department of Astronomy, University of California, Berkeley, CA 94720, USA}

\author[0000-0002-1033-3656]{M.~A.~Wozniak}
\affil{Department of Physics \& Astronomy, University of California Riverside, 900 University Ave, Riverside, CA}

\author[0000-0002-0840-6940]{S.~K.~Yadavalli}
\affiliation{Department of Astronomy and Astrophysics, Pennsylvania State University, 525 Davey Laboratory, University Park, PA 16802, USA}

\author[0000-0002-0632-8897]{Y.~Zenati}
\affiliation{Department of Physics and Astronomy, The Johns Hopkins University, Baltimore, MD 21218, USA}
\affil{ISEF International Fellowship}

\begin{abstract}
We present UV/optical observations and models of supernova (SN) 2023ixf, a type II SN located in Messier 101 at 6.9~Mpc. Early-time (``flash'') spectroscopy of \sn{}, obtained primarily at Lick Observatory, reveals emission lines of \ion{H}{i}, \ion{He}{i/ii}, \ion{C}{iv}, and \ion{N}{iii/iv/v} with a narrow core and broad, symmetric wings arising from the photo-ionization of dense, close-in circumstellar material (CSM) located around the progenitor star prior to shock breakout. These electron-scattering broadened line profiles persist for $\sim$8~days with respect to first light, at which time Doppler broadened features from the fastest SN ejecta form, suggesting a reduction in CSM density at $r \gtrsim 10^{15}$~cm.  The early-time light curve of \sn{} shows peak absolute magnitudes (e.g., $M_{u} = -18.6$~mag, $M_{g} = -18.4$~mag) that are $\gtrsim 2$~mag brighter than typical type II supernovae, this photometric boost also being consistent with the shock power supplied from CSM interaction. Comparison of \sn{} to a grid of light curve and multi-epoch spectral models from the non-LTE radiative transfer code \cmfgen\ and the radiation-hydrodynamics code \heracles\ suggests dense, solar-metallicity, CSM confined to $r = (0.5-1) \times 10^{15}$~cm and a progenitor mass-loss rate of $\dot{M} = 10^{-2}~\Msun$~yr$^{-1}$. For the assumed progenitor wind velocity of $v_w = 50~\kms$, this corresponds to enhanced mass-loss (i.e., ``super-wind'' phase) during the last $\sim$3-6~years before explosion. 


\end{abstract}


\section{Introduction} \label{sec:intro}

A paramount issue in astrophysics is constraining how the lives of red supergiant (RSG) stars end. This avenue of study has a direct impact on the observed diversity of core-collapse supernovae (SNe), compact object formation, and element creation in the Universe. Advancing our understanding of late-stage RSG evolution can be accomplished by probing the composition and structure of the circumstellar medium (CSM) surrounding these stars in the final years before explosion. This CSM is composed of material once located on the RSG surface and is enriched as the progenitor star loses mass via winds and/or violent outbursts \citep{smith14}. Understanding the structure of this CSM provides needed constraints on the final stages of stellar evolution before core collapse and the proposed mechanisms for both dynamic (e.g., gravity waves, super-Eddington winds; \citealt{owocki04, owocki17, Quataert12, quataert16, fuller17,Wu21}) as well as secular (i.e., steady-state wind; \citealt{Beasor20}) mass-loss.


Early-time, multi-wavelength observations of young type II SNe (SNe II) are an essential probe of the final stages of stellar evolution; these last months-to-centuries are almost completely unconstrained in stellar evolutionary models. In the era of all-sky transient surveys, ``flash'' or rapid spectroscopic observations have become a powerful tool in understanding the direct circumstellar environment of SN progenitors in the final months before explosion (e.g., \citealt{galyam14, Groh14, smith15, Khazov16, Yaron17, wjg20, bruch21, bruch22, wjg22, terreran22, Tinyanont22, davis23, Wang23}). Very early-time spectra ($t \lesssim 7$~days) can be used to identify prominent emission lines from the recombination of CSM photo-ionized by the incoming SN radiation at, and following, shock breakout. However, these spectral features are transient, leaving behind broad lines from the fastest ($v_{\rm ej} \approx 10^{4}$~km~s$^{-1}$) SN ejecta layers \citep{Chugai01,dessart17}. The strength of the narrow emission features depends on the CSM density and its chemical abundance. This is a robust tracer of the progenitor's chemical composition, identity and recent mass-loss at small distances \citep[$r < 10^{15}$~cm from the explosion;][]{galyam14-1, Yaron17, dessart17, Boian20}. 



Spectroscopic modeling of CSM-interacting SNe~II with non-LTE radiative transfer codes (e.g., \cmfgen; \citealt{hillier12, D15_2n}) has been used to extract quantitative information on the progenitor's radius, chemical composition, wind velocity and mass-loss rate. \cmfgen\ in particular allows for self-consistent post-processing of radiation hydrodynamics simulations, allowing for physically robust constraints on both the SN ejecta and CSM properties (shocked and unshocked) and the creation of accurate synthetic spectra that contain critical information absent from light curves. For example, \cmfgen\ spectral modeling of the prototypical ``flash'' spectroscopy type II SN 1998S indicated an enhanced RSG mass-loss rate of $\dot{M} \approx (0.6-1) \times 10^{-2}$~M$_{\odot}$~yr$^{-1}$ ($v_w \approx 50~\kms$) in the final 15 years before core-collapse \citep{Shivvers15, Dessart16}. This mass-loss rate is significantly larger than that observed in galactic RSGs e.g., $\dot{M} \approx 10^{-6}$~M$_{\odot}$~yr$^{-1}$ \citep{Beasor20}.  Similarly, modeling of the emission line spectrum in the first days of SNe~2017ahn and 2020pni \citep{Tartaglia21, terreran22} suggested N-rich CSM derived from mass-loss rates of $\dot{M} \approx (3-6) \times 10^{-3}$~M$_{\odot}$~yr$^{-1}$. Further diversity was revealed from \cmfgen\ modeling of SN~2013fs \citep{Yaron17}, observed within hours of explosion, which suggested a compact CSM ($r < 3\times 10^{14}$~cm) with a lower mass-loss rate of $\dot{M} \approx (3-5)\times 10^{-3}$~M$_{\odot}$~yr$^{-1}$ \citep{dessart17}. Additionally, both UV/optical photometry and the spectral series of SN~2020tlf were accurately modeled by \cmfgen\ simulations involving an inflated RSG progenitor ($R_{\star} \approx 10^3~\Rsun$, which exploded into a dense ($\dot{M} = (1-3) \times 10^{-2}$~M$_{\odot}$~yr$^{-1}$, $v_w = 50~\kms$), extended ($r \approx 10^{15}$~cm) CSM; this scenario is also consistent with the detection of luminous precursor emission before explosion \citep{wjg22}.



In this paper we present, analyze, and model photometric and spectroscopic observations of \sn{}, first reported to the Transient Name Server by Koichi Itagaki \citep{Itagaki23} on 2023-05-19 (MJD 60083.90). \sn{} was classifed as a Type II SN \citep{perley23} and is located at $\alpha = 14^{\textrm{h}}03^{\textrm{m}}38.56^{\textrm{s}}$, $\delta = +54^{\circ}18'41.96^{\prime \prime}$ in host galaxy Messier 101 (NGC 5457). Based on reported pre-discovery images from numerous observations of \sn{}, we adopt a time of first light to be MJD $60082.833 \pm 0.020$ \citep{Mao23} that is based on the average between last non-detection and first detection, but could be earlier given the shallow depth of the last non-detection limit. All phases reported in this paper are with respect to this adopted time of first light. In this paper, we use a redshift-independent host-galaxy distance of $6.85 \pm 0.15$~Mpc reported by \cite{riess22}, which is the updated value beyond what is presented in \cite{riess16}. We adopt a redshift of $z = 0.000804$ \citep{perley23}.

Given its close proximity and current relative brightness, \sn{} represents an unparalleled opportunity to study both the very early-time and the long-term evolution of a CSM-interacting SN~II in unprecedented detail. In \S\ref{sec:obs} we describe UV, optical, and NIR observations of SN~2023ixf. In \S\ref{sec:analysis} we present analysis, comparisons and modeling of \sn{}'s optical photometric and spectroscopic properties. Finally, in \S\ref{sec:discussion} we discuss the progenitor environment and mass-loss history prior to \sn{}. Conclusions are drawn in \S\ref{sec:conclusion}. All uncertainties are quoted at the 68\% confidence level (c.l.) unless otherwise stated.


\section{Observations} \label{sec:obs}


\subsection{Photometric Observations}\label{SubSec:Phot}

\sn{} was observed with the Pan-STARRS telescope \citep[PS1/2;][]{Kaiser2002, Chambers2017} between 21 May and 2 June 2023 in $grizy$-bands through the Young Supernova Experiment (YSE) \citep{Jones2021}. Data storage/visualization and follow-up coordination was done through the YSE-PZ web broker \citep{Coulter22, Coulter23}. The YSE photometric pipeline is based on {\tt photpipe} \citep{Rest+05}, which relies on calibrations from \citep{Magnier20a, waters20}. Each image template was taken from stacked PS1 exposures, with most of the input data from the PS1 3$\pi$ survey. All images and templates were resampled and astrometrically aligned to match a skycell in the PS1 sky tessellation. An image zero-point is determined by comparing PSF photometry of the stars to updated stellar catalogs of PS1 observations \citep{flewelling16}. The PS1 templates are convolved with a three-Gaussian kernel to match the PSF of the nightly images, and the convolved templates are subtracted from the nightly images with {\tt HOTPANTS} \citep{becker15}. Finally, a flux-weighted centroid is found for the position of the SN in each image and PSF photometry is performed using ``forced photometry": the centroid of the PSF is forced to be at the SN position. The nightly zero-point is applied to the photometry to determine the brightness of the SN for that epoch.

We obtained $ugri$ imaging of SN\,2023ixf with the Las Cumbres Observatory (LCO) 1\,m telescopes from 20 May to 1 June 2023 (Programs NSF2023A-011 and NSF2023A-015; PIs Foley and Kilpatrick, respectively).  After downloading the {\tt BANZAI}-reduced images from the LCO data archive \citep{mccully18}, we used {\tt photpipe} \citep{Rest+05} to perform {\tt DoPhot} PSF photometry \citep{Schechter+93}. All photometry was calibrated using PS1 stellar catalogs described above with additional transformations to SDSS $u$-band derived from \citet{finkbeiner16}. For additional details on our reductions, see \citet{kilpatrick18}. We also obtained photometry using a 0.7 meter Thai Robotic Telescope at Sierra Remote Observatories and the Nickel Telescope at Lick Observatory in the $BVRI$ bands. Images are bias subtracted and field flattened. Absolute photometry is obtained using stars in the 10$'\times$10$'$ field of view.

\begin{figure*}[t]
\centering
\includegraphics[width=0.99\textwidth]{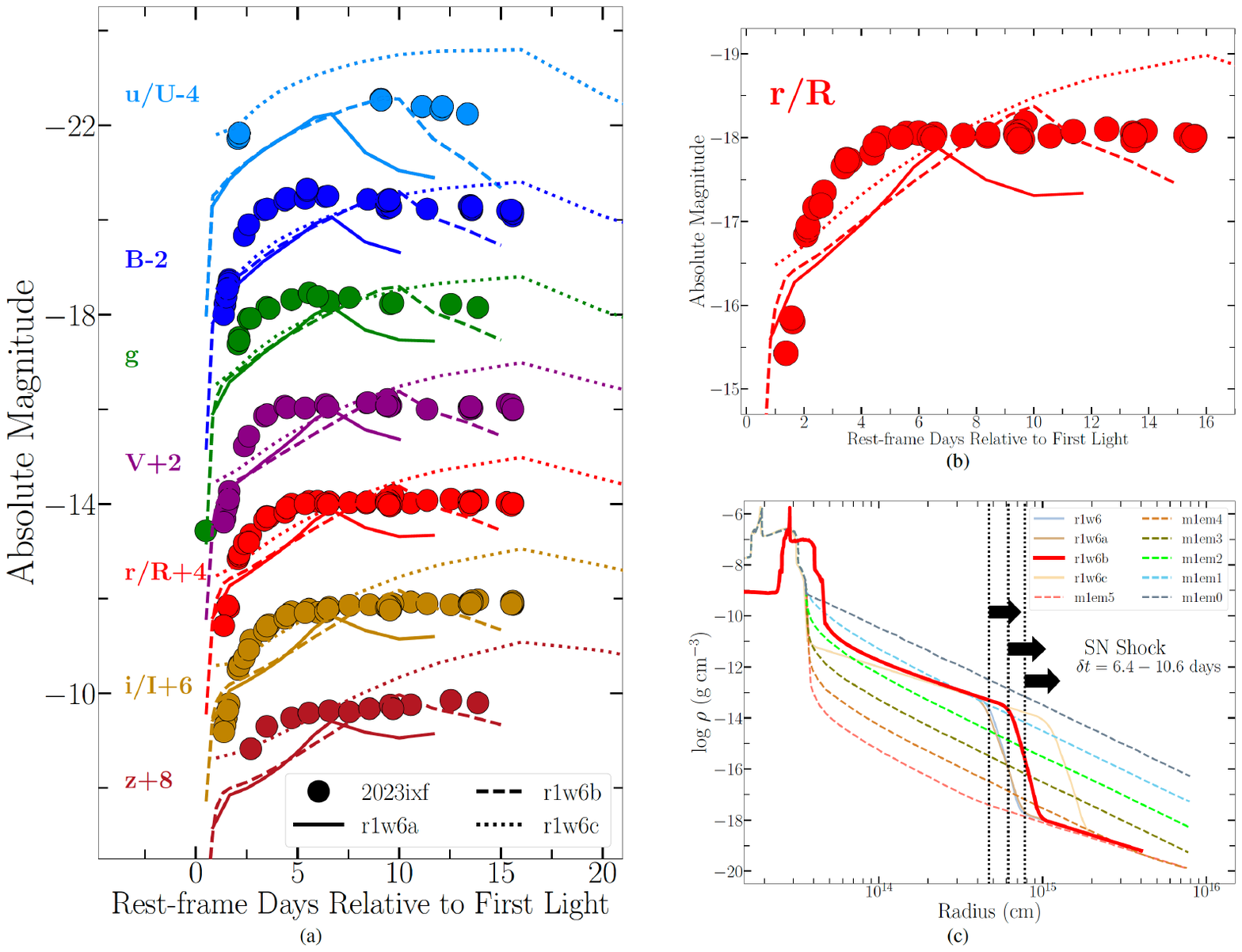}
\caption{(a) Multi-color light curve of \sn{} (circles) with respect to time since first light (MJD $60082.833 \pm 0.020$) from PS1, LCO, Auburn, TRT, Nickel, and Lulin telescopes. Observed photometry is presented in the AB magnitude system and has been corrected for host galaxy and MW extinction. Light curves of \cmfgen\ models r1w6a ($R_{\rm CSM} = 6\times 10^{14}$~cm, $\dot{M} = 10^{-2}~\Msun$~yr$^{-1}$), r1w6b ($R_{\rm CSM} = 8\times 10^{14}$~cm, $\dot{M} = 10^{-2}~\Msun$~yr$^{-1}$), and r1w6c ($R_{\rm CSM} = 1\times 10^{15}$~cm, $\dot{M} = 10^{-2}~\Msun$~yr$^{-1}$) are plotted as solid, dashed and dotted lines, respectively. (b) Zoom-in of the \sn{} $r/R-$band light curve and \cmfgen\ models, which can reproduce the peak magnitude but are inconsistent with the early-time slope. (c) CSM densities and radii for a subset of the \cmfgen\ model grid (e.g., Table \ref{tab:models}) used to find the best fitting model for \sn{}, which is plotted as a solid red line (r1w6b). Dotted black lines represent lower limits on the location of the SN shock at $\delta t = 6.4, 8.4, 10.6$~days, for a lower limit on the SN shock velocity of $\gtrsim 8500~\kms$ (\S\ref{subsec:spec_analysis}). We expect a decrease of the optical depth to electron-scattering (i.e., $\tau_{\rm ES}$) based on the plotted density profile at around $\sim$8~days, which is consistent with the fading of the IIn-like line profiles observed in \sn{} at these phases. 
\label{fig:LC_rho}}
\end{figure*}

We also observed SN\,2023ixf with the Lulin 1\,m telescope in $griz$ bands from 21 May to 1 June 2023.  Standard calibrations for bias and flat-fielding were performed on the images using {\tt IRAF}, and we reduced the calibrated frames in {\tt photpipe} using the same methods described above for the LCO images.

We also observed SN\,2023ixf with the Auburn 10'' telescope located in Auburn, AL from 27 May to 3 June 2023 in $BGR$ bands.  Following standard procedures in {\tt python}, we corrected each frame for bias, dark current, and flat-fielding using image frames obtained in the same instrumental setup.  We then registered each frame using {\it Gaia} DR3 astrometric standard stars \citep{GaiaDR3cat} observed in the same field as each image.  Finally, we stacked images in each filter for each night with {\tt swarp} and performed final photometry using {\tt DoPhot} with calibration using Pan-STARRS $gri$ standard stars transformed to $BVR$ bands\footnote{Note that our $G$-band filter is close to Johnson $V$-band, and so we calibrate against Pan-STARRS standard stars transformed into this band.  For filter functions, see \url{https://astronomy-imaging-camera.com/product/zwo-lrgb-31mm-filters-2}.}. The complete multi-color light curve of \sn{} is presented in Figure \ref{fig:LC_rho}(a)). 

The Milky Way (MW) $V$-band extinction and color excess along the SN line of site is $A_{V} = 0.025$~mag and \textit{E(B-V)} = 0.008~mag \citep{schlegel98, schlafly11}, respectively, which we correct for using a standard \cite{fitzpatrick99} reddening law (\textit{$R_V$} = 3.1). In addition to the MW color excess, we estimate the contribution of galaxy extinction in the local SN environment. Using a high resolution Kast spectrum of \sn{} at $\delta t = 2.4$~days, we calculate \ion{Na}{i} D2 and D1 equivalent widths (EWs) of 0.16 and 0.12~\AA, respectively; these values are consistent with those derived from a Keck Planet Finder spectrum (Lundquist et al. 2023). We use Equations 7 \& 8 in \cite{Poznanski12} to convert these EWs to an intrinsic \textit{E(B-V)} and find a host galaxy extinction of $E(B-V)_{\textrm{host}} = 0.033 \pm 0.010$~mag, also corrected for using the \cite{fitzpatrick99} reddening law.

\subsection{Spectroscopic Observations}\label{SubSec:Spec}

\begin{figure*}[t]
\centering
\includegraphics[width=0.99\textwidth]{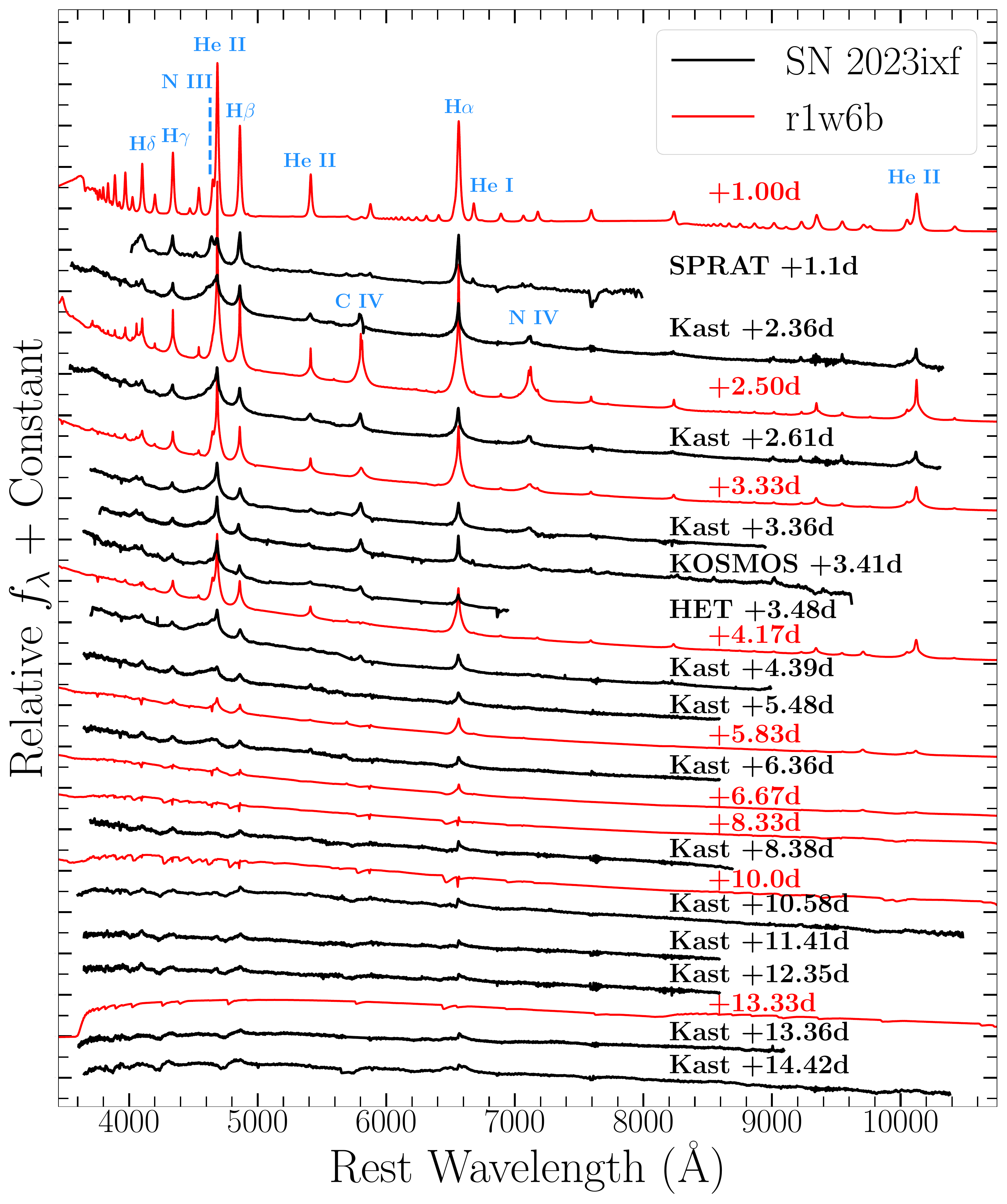}
\caption{Early-time spectral series of \sn{} (black) with respect to \cmfgen\ model r1w6b (red), which is characterized by a wind mass-loss rate of $\dot{M} = 10^{-2}~\Msun$~yr$^{-1}$ that extends to a CSM radius of $R_{\rm CSM} = 8\times 10^{14}$~cm. Model spectra have been smoothed with a Gaussian kernel to match the spectral resolution of the data. Line identifications shown in blue. The disappearance of \ion{He}{i} and \ion{N}{iii} after the $\delta t = +1.1$~day spectrum indicates a rise in ionization and temperature in \sn{} following the propagation of the shock breakout radiation. \label{fig:spec_series} }
\end{figure*}

\sn{} was observed with Shane/Kast \citep{KAST} between $\delta t = 2.4 - 14.4$~days. For all these spectroscopic observations, standard CCD processing and spectrum extraction were accomplished with \textsc{IRAF}\footnote{\url{https://github.com/msiebert1/UCSC\_spectral\_pipeline}}. The data were extracted using the optimal algorithm of \citet{1986PASP...98..609H}.  Low-order polynomial fits to calibration-lamp spectra were used to establish the wavelength scale and small adjustments derived from night-sky lines in the object frames were applied. \sn{} spectra were also obtained with the Kitt Peak Ohio State Multi-Object Spectrograph (KOSMOS, \citealt{KOSMOS}) on the Astrophysical Research Consortium (ARC) 3.5-meter Telescope at Apache Point Observatory (APO). The KOSMOS spectra were reduced through the {\tt KOSMOS}\footnote{\url{https://github.com/jradavenport/pykosmos}} pipeline. One optical spectrum (in a red and blue arm) was taken through the Low-Resolution Spectrograph 2 (LRS2) instrument on the Hobby Eberly Telescope (HET) on 2023-05-21 (blue arm) and 2023-05-22 (red arm). The LRS2 data were processed with \texttt{Panacea}\footnote{\url{https://github.com/grzeimann/Panacea}}, the HET automated reduction pipeline for LRS2.  The initial processing includes bias correction, wavelength calibration, fiber-trace evaluation, fiber normalization, and fiber extraction; moreover, there is an initial flux calibration from default response curves, an estimation of the mirror illumination, as well as the exposure throughput from guider images.  After the initial reduction, we use LRS2Multi\footnote{\url{https://github.com/grzeimann/LRS2Multi}} in order to perform sky subtraction.


In Figure \ref{fig:spec_series} we present the complete series of optical spectroscopic observations of \sn{} from $\delta t = 2.4-14.4$~days. In this plot, we also show the classification spectrum of \sn{} at $+1.1$~days from the Liverpool telescope \citep{perley23}. However, because we cannot verify the quality of this spectral reduction, we only use these data for narrow line identification. Additionally, we include {\it Swift} UV grism spectra of \sn{} from $\delta t = +1.8 - 2.8$~days in Appendix Figure \ref{fig:spec_UV}; data were reduced using the techniques outlined in \cite{Pan20}. The complete spectral sequence is shown in Figure \ref{fig:spec_series} and the log of spectroscopic observations is presented in Appendix Tables \ref{tab:spec_table}.

\begin{figure*}
\centering
\subfigure[]{\includegraphics[width=0.33\textwidth]{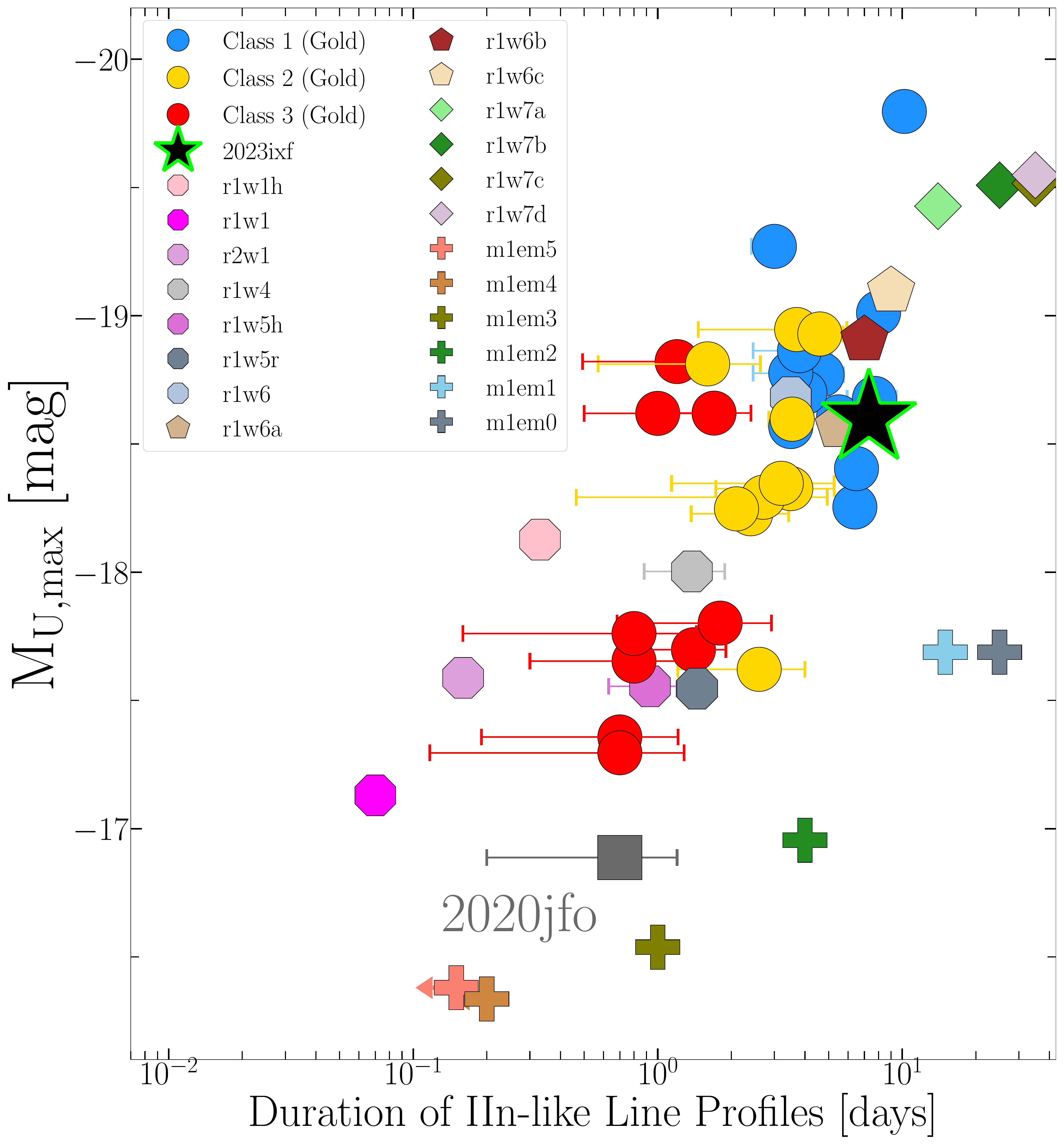}}
\subfigure[]{\includegraphics[width=0.33\textwidth]{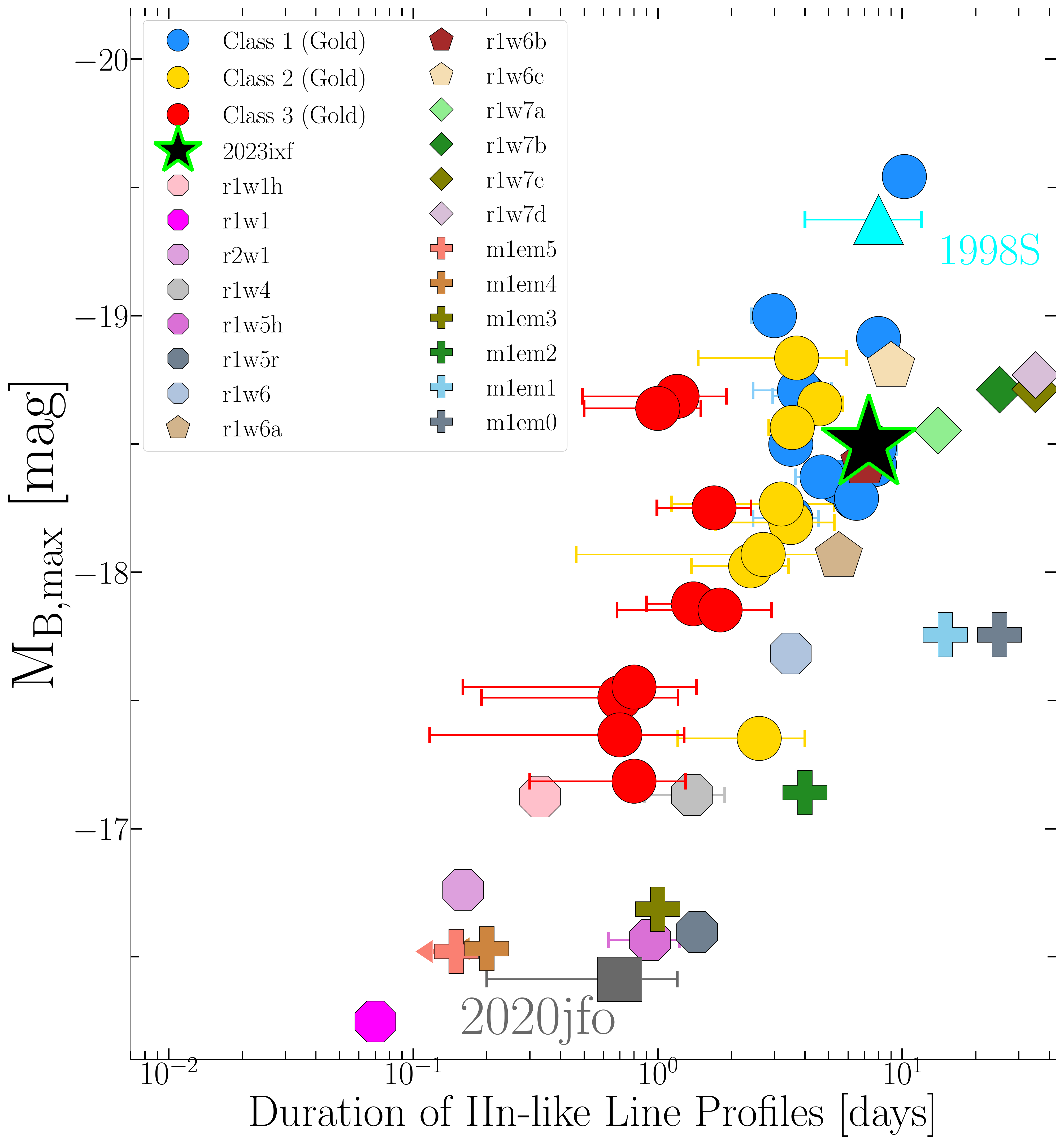}}
\subfigure[]{\includegraphics[width=0.33\textwidth]{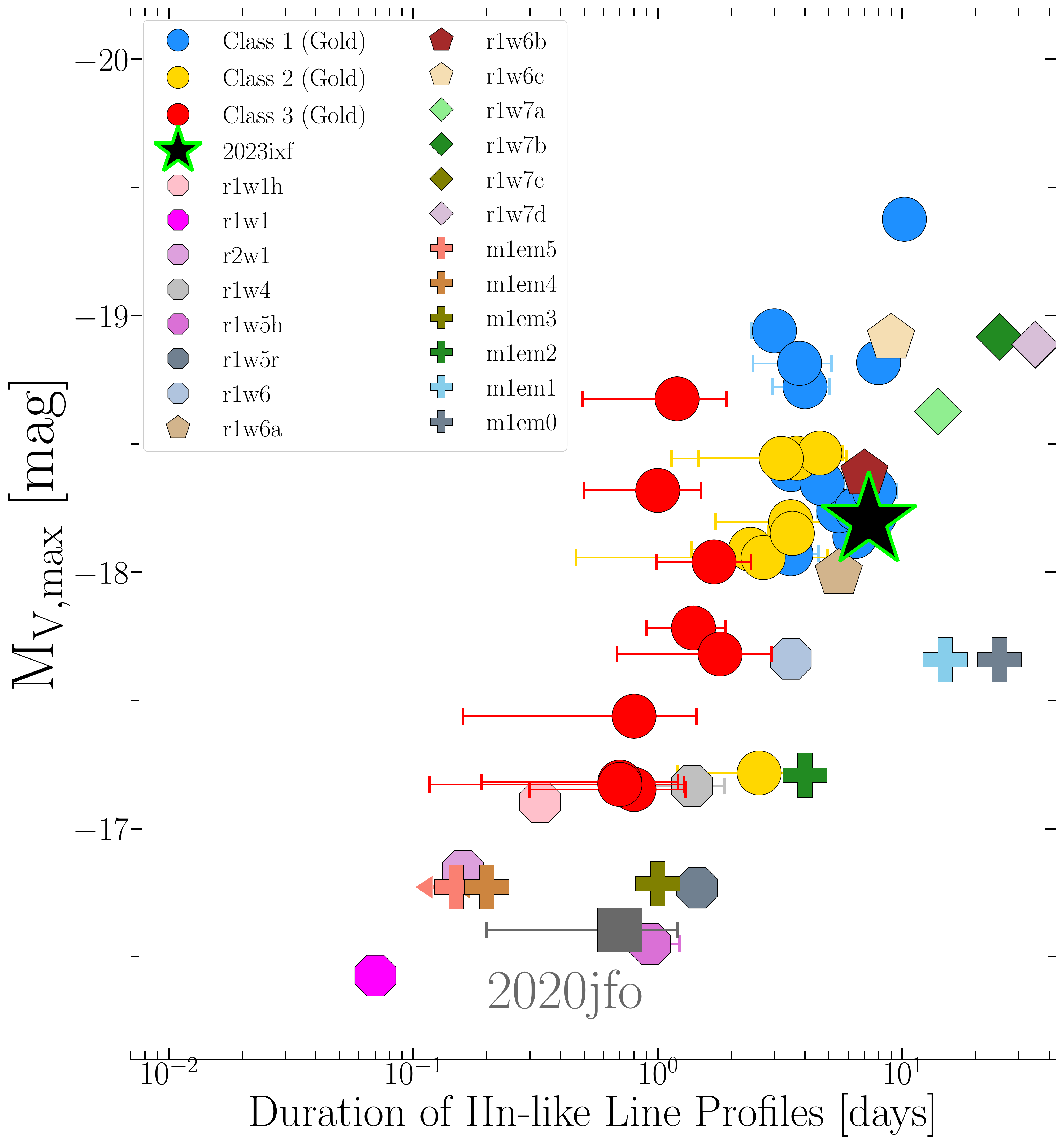}}
\caption{Peak (a) $U-$band, (b) $B-$band, and (c) $V-$band absolute magnitudes versus duration of IIn-like line profiles ($t_{\rm IIn}$) for \sn{} (black star) with respect to CSM-interacting SNe~II presented in \cite{Jacobson-Galan23} shown as circles (color delineation discussed in \S\ref{subsec:spec_analysis}). SN~2020jfo \citep{Teja22} shown as reference canonical SN~II without significant CSM interaction (grey square). \cmfgen\ models plotted as colored octagons, polygons, diamonds and plus signs. \sn{} has an observed $t_{\rm IIn}$ and peak absolute magnitude that is consistent with the other gold sample SNe~II displaying the strongest signs of CSM interaction e.g., SNe~1998S, 2017ahn, 2020pni, 2020tlf \citep{Leonard00,fassia01, Shivvers15, Tartaglia21, terreran22, Jacobson-Galan23} 
\label{fig:max_lines} }
\end{figure*}

\section{Analysis}\label{sec:analysis}

\subsection{Photometric Properties}\label{subsec:phot_properties}

The complete early-time, multi-band light curve of \sn{} is presented in Figure \ref{fig:LC_rho}(a). We fit a $5^{\rm th}$-order polynomial to the $g-$band light curve to derive a peak absolute $g-$band magnitude of $M_g = -18.4 \pm 0.10$\,mag at MJD $60088.61\pm0.10$, where the uncertainty on peak magnitude is the $1\sigma$ error from the fit and the uncertainty on the peak phase is found from adding the uncertainties on both the time of peak magnitude and the time of first light in quadrature. Using the adopted time of first light, this indicates a rise time of $t_r = 5.8 \pm 0.10$\,days with respect to $g$-band maximum. Other filters display similarly bright peak absolute magnitudes e.g., $M_u = -18.6 \pm 0.11$\,mag and $M_r = -18.0 \pm 0.09$\,mag --- this indicates a bolometric boost to the SN brightness rather than a color effect. Following its rise to peak, the multi-color light curve of \sn{} has remained at an approximately constant brightness, indicating that it could be entering a plateau phase (i.e., SN~II-P classification). However, at this time, the SN is still very blue, therefore indicating that the recombination phase has yet to be reached. All peak magnitudes and rise-times are presented in Table \ref{table:props}.


In Figure \ref{fig:max_lines}, we compare the observed peak absolute magnitudes of \sn{} to a sample of SNe~II with spectroscopic signatures of CSM-interaction (i.e., IIn-like profiles) \citep{Jacobson-Galan23}. This gold sample includes most of the known CSM-interacting SNe~II that show detectable IIn-like profiles in their early-time spectra and have early-time UV observations with {\it Swift} UVOT. The color delineation of all presented sample objects is as follows: at phases of $t\sim 2$~days post-explosion, blue colored objects show high ionization emission lines of \ion{N}{iii}, \ion{He}{ii}, and \ion{C}{iv} (e.g., SNe~1998S, 2017ahn, 2018zd, 2020pni, 2020tlf, etc), yellow colored objects have no \ion{N}{iii} emission but do show \ion{He}{ii}, and \ion{C}{iv} (e.g., SNe~2014G, 2022jox), and red colored objects only show weaker \ion{He}{ii} emission (e.g., SNe~2013fs). However, it should be noted that high ionization lines of \ion{O}{v/vi}, \ion{C}{v}, and \ion{N}{iv} are also present in SN~2013fs at $t < 1$~day due to a more compact CSM than other CSM-interacting SNe~II \citep{Yaron17, dessart17}. With respect to other SNe~II with evidence for interaction with CSM, \sn{} is $\sim$0.5~mag brighter in all observed filters than the median peak absolute magnitude observed in the sample. \sn{} has a comparable peak brightness and rise-times to SNe~2017ahn, 2018zd, 2020pni, 2020abjq and 2022ffg \citep{zhang20, Hiramatsu21, Tartaglia21, terreran22, Jacobson-Galan23}, all of which have similar early-time spectral morphology and duration of the IIn-like line profiles (\S\ref{subsec:spec_analysis}). However, the rise-time of \sn{} is significantly shorter than more extreme events such as SNe~1998S, 2019qch, 2020tlf, 2021tyw and 2022pgf, whose rise-times are $>$12~days. This difference reflects a shorter interaction timescale in a less extended, high-density CSM in \sn{}. Furthermore, \sn{} is distinct from other prototypical SNe~II with IIn-like profiles such as SN~2013fs and 2014G \citep{terreran16, Yaron17}, which have shorter rise-times and lower peak absolute magnitudes. Finally, \sn{} is $\sim$2~mags brighter in multi-band (i.e., $uBVgriz$) filters than SNe~II without IIn-like profiles in their early-time spectra e.g., SN~2020jfo (\citealt{Teja22}; Fig. \ref{fig:max_lines}) or average values derived from samples of SNe~II \citep{anderson14}.   

\subsection{Spectroscopic Properties}\label{subsec:spec_analysis}

The complete early-time spectroscopic sequence of \sn{} from $\delta t = +1.1$ to $+14.4$~days is presented in Figure \ref{fig:spec_series}, consistent with other spectral sequences released on this object \citep{Stritzinger23, Yamanaka23}.  In the earliest spectrum, \sn{} shows narrow, symmetric emission features of \ion{H}{i}, \ion{He}{i/ii}, \ion{N}{iii/iv} and \ion{C}{iv}. A two-component Lorentzian model fit to the H$\alpha$ profile in the high resolution ($R\approx3000$) +2.4~d Kast spectrum shows a narrow component full width at half maximum (FWHM) velocity of $<$150~$\kms$ and broad symmetric component velocity of $\sim$1400~$\kms$; the former is ascribed to the progenitor wind while the latter is caused by scattering of recombination line photons by free, thermal electrons in the ionized CSM \citep{Chugai01, Dessart09, Huang18}. However, it should be noted that at these phases, there could be radiative acceleration of the CSM that causes the width of the narrow component to be larger than the true velocity of the progenitor wind \citep{D15_2n,Tsuna23}.



\sn{} may be the first SN to exhibit a rapid rise in ionization between the first and second spectrum as shown in Figure \ref{fig:spec_series}. This is caused initially by the shock breakout pulse and later on by the incoming radiation from the embedded shock. This is witnessed here with the presence of lines of moderately ionized species (i.e. lines of \ion{He}{i} or \ion{N}{iii}) and a moderately blue color at $\delta t = 1.1$~days. At $\delta t = 2.4$~days, the \sn{} spectrum is much bluer, the lines of \ion{He}{i}~$\lambda\lambda\lambda$5875, 6678, and 7065 have weakened or disappeared, and the spectrum exhibits instead lines of \ion{C}{iv} ($\lambda$5808) and \ion{He}{ii} ($\lambda$4686). Furthermore, there is emission from \ion{N}{v}~$\lambda4604-4620$ contributing at wavelengths bluewards of the \ion{He}{ii}~$\lambda 4686$ line, consistent with heightened ionization at these phases. 

\begin{figure*}
\centering
\subfigure[]{\includegraphics[width=0.33\textwidth]{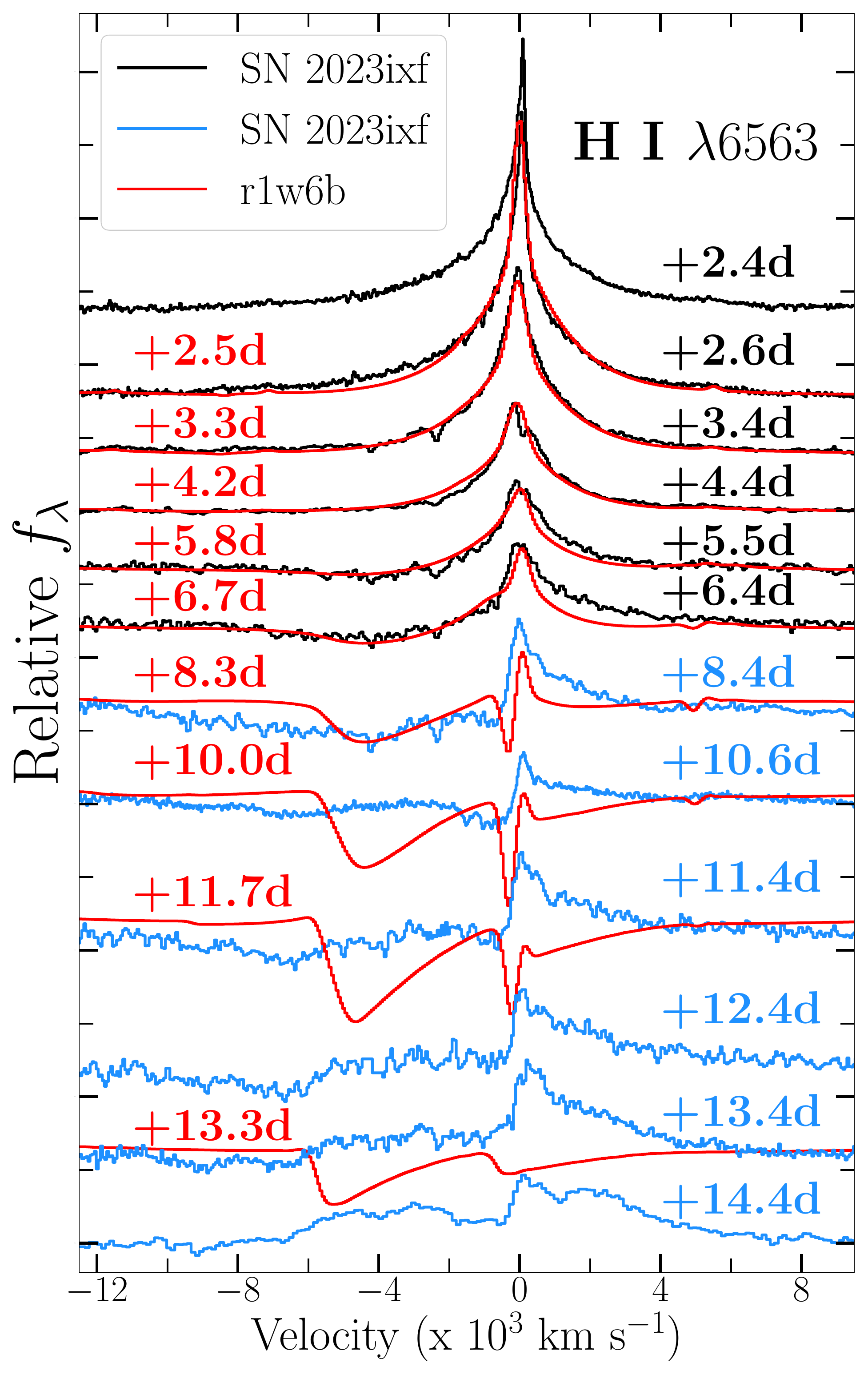}}
\subfigure[]{\includegraphics[width=0.33\textwidth]{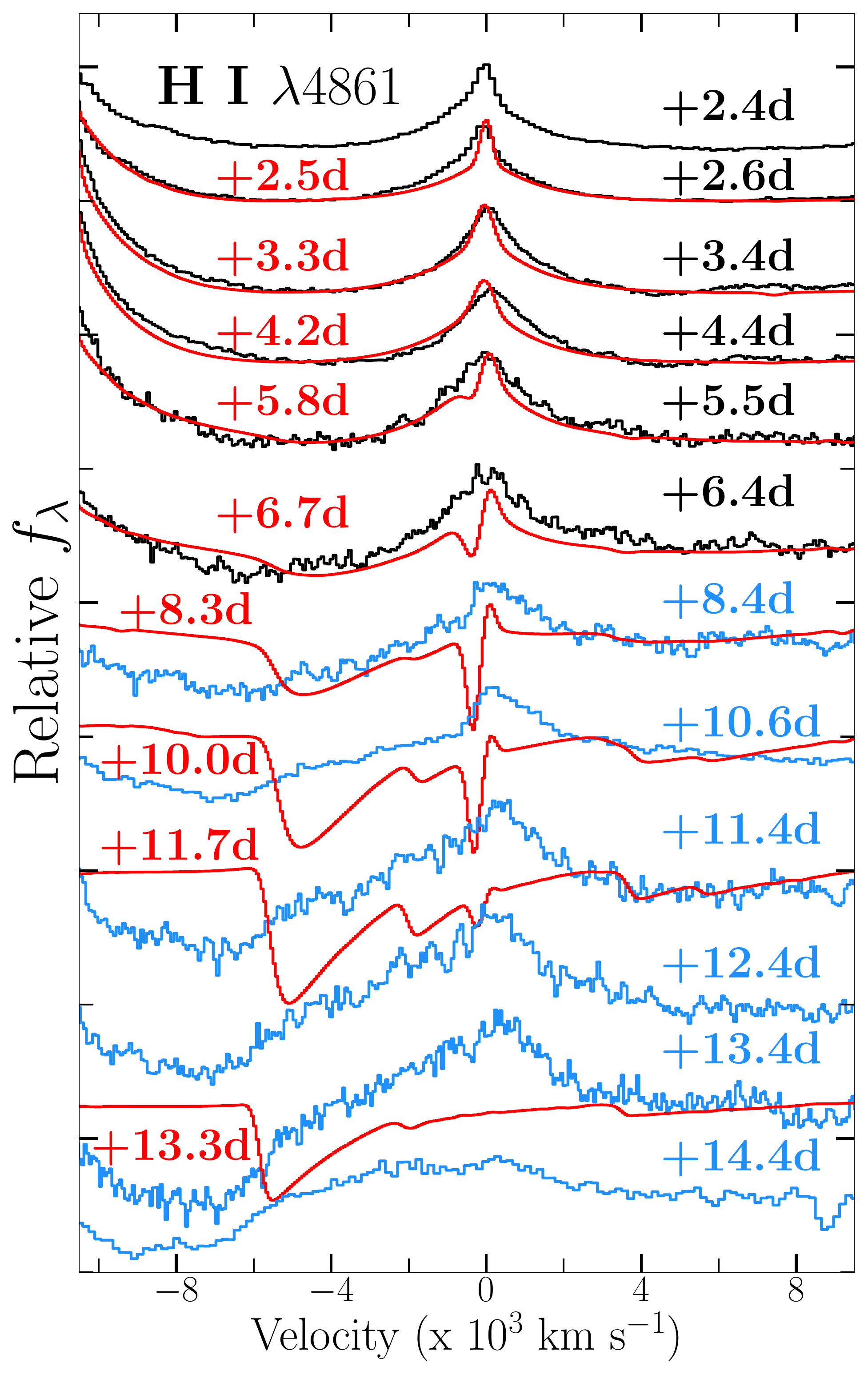}}
\subfigure[]{\includegraphics[width=0.33\textwidth]{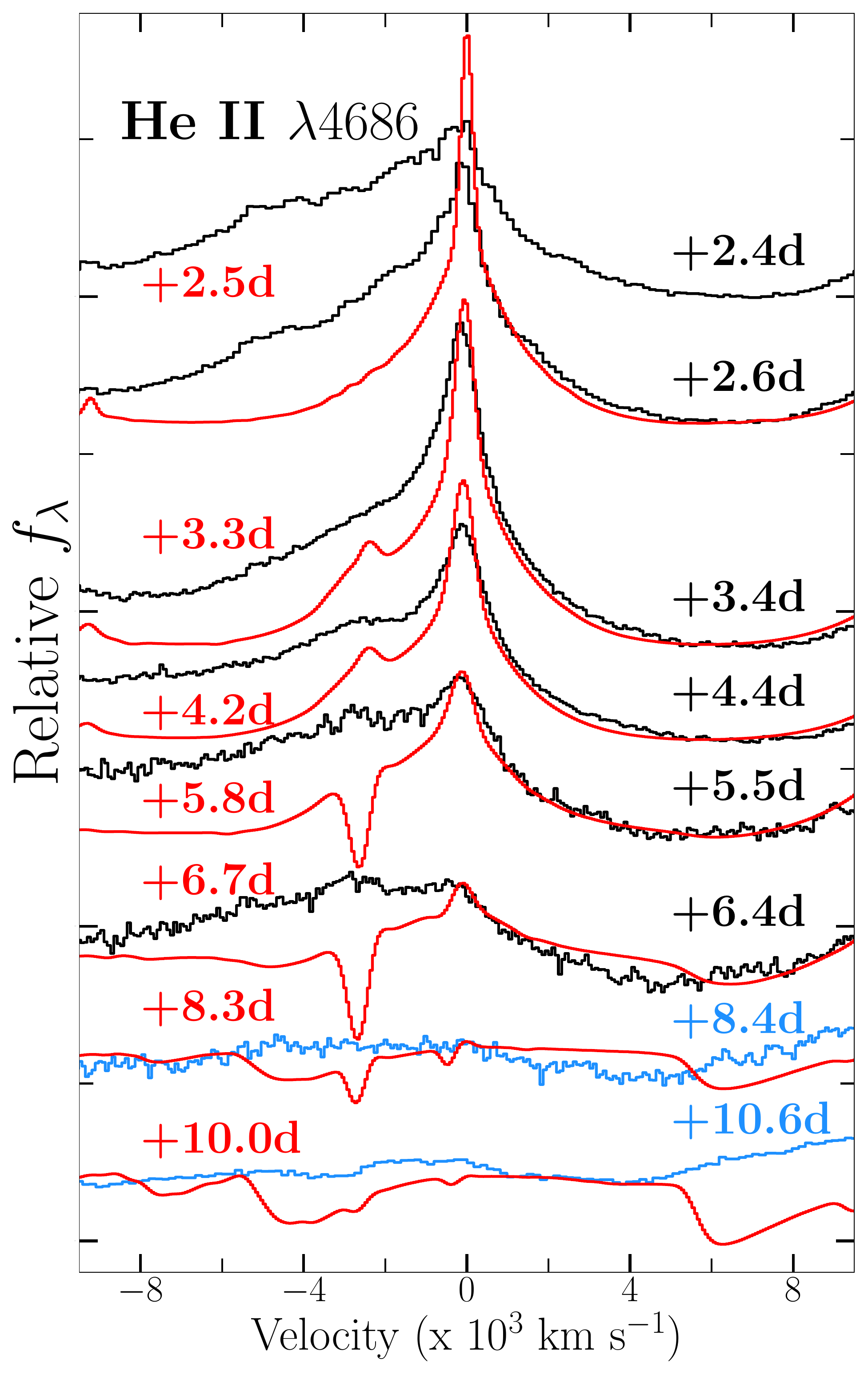}}
\caption{(a) H$\alpha$ velocity evolution in \sn{} from $\delta t = 2.4 - 14.4$~days with respect to r1w6b model spectra (red), which has been scaled to the emission line peaks of \sn{} and smoothed with a Gaussian filter to better compare with the data. Early-time spectral profiles are shaped by electron-scattering in the dense CSM. The transition shown from black to blue lines ($t_{\rm IIn} \approx 8$~days) marks the emergence of broad absorption features derived from the fastest moving SN ejecta. (b) H$\beta$ velocity evolution, also showing that the electron-scattering line profiles subside after $\sim 8$~days. (c) \ion{He}{ii}~$\lambda$4686 velocity evolution reveals that the electron-scattering profile fades by $\sim 8$~days, suggesting a significant decrease in CSM density.   
\label{fig:vels} }
\end{figure*}

The narrow, symmetric line profiles with Lorentzian wings caused by electron-scattering (i.e., IIn-like) continue to persist in \sn{} for the first week of the SN evolution. Then, in the +5.48~d and +6.36~d spectra, the \ion{He}{ii} emission begins to fade (Fig. \ref{fig:vels}) and the SN develops a broad absorption profile in all Balmer transitions, indicating the escape of photons from the fast moving ejecta and a decrease in CSM density. We therefore define the duration of the IIn-like line profiles as the transition point at which the optical depth to electron-scattering has dropped enough to see the emerging fast-moving SN ejecta. For \sn{}, we estimate that this change occurs at $t_{\rm IIn} \approx 8 $~days, which is reflective of the disappearance of the electron-scattering wings in the \ion{He}{ii} emission line and the development of broad absorption profiles at Balmer series wavelengths. This indicates that the photosphere is first located in the unshocked CSM (far above the shock), then in the swept up material present in the fast moving dense shell (i.e., shocked CSM), and then in the fastest moving SN ejecta below the dense shell. Based on the $T > 10$~keV X-ray spectrum of \sn{} \citep{Grefenstette23}, there is sufficiently high temperatures for \ion{He}{ii} to exist, so the decrease in line strength can be attributed to a reduction in particle density as the shock samples CSM at $r>10^{15}$~cm. As shown in Figure \ref{fig:vels}, the bluest edge of the H$\alpha$ and H$\beta$ line profiles corresponds to velocities of $\sim$8500~$\kms$, which provides a lower limit on the velocities of the fastest moving H-rich material at the shock front. By two weeks post-explosion, the SN spectra is composed of broad \ion{H}{i} absorption profiles, similar to other young SNe~II. 

The duration of the IIn-like signatures in \sn{} is consistent with other CSM-interacting SNe~II with enhanced progenitor mass-loss rates of $\dot{M} > 10^{-2}~\Msun$~yr$^{-1}$ (Figure \ref{fig:max_lines}). In Figure \ref{fig:max_lines}, we present peak absolute magnitudes with respect to IIn profile duration for all gold sample CSM-interacting SNe~II analyzed in \cite{Jacobson-Galan23}.  Intriguingly, there exists a natural trend between peak brightness and IIn-like profile duration amongst these events, which is reflective of a continuum of progenitor mass-loss rate and CSM extent. The observed $t_{\rm IIn}$ in \sn{} is most similar to SNe 2017ahn, 2018zd, 2020pni, 2020tlf and 2022ffg, but is not as large as that observed in 2020tlf, 2021tyw, or 2022pgf, likely due to a more extended dense CSM in those objects. Furthermore, the evolution of \sn{} is unlike other CSM-interacting SNe~II with $t_{\rm IIn} < 5$~days post-explosion (e.g., SNe~2013fs or 2014G), which do not show N emission lines at phases $>1$~day post-explosion and is caused by a more compact/high density or extended/low density CSM \citep{dessart17}. The observed early spectral differences between \sn{} and other CSM-interacting SNe~II is illustrated in Figure \ref{fig:spec_compare}. Here it is shown that the line profiles of \sn{} at $\sim2$~days post-explosion are most consistent with interaction between SN ejecta and CSM constructed from a high progenitor mass-loss rate ($\dot{M} > 10^{-2}~\Msun$~yr$^{-1}$; also see Fig. \ref{fig:model_all}). 


\subsection{Modeling}\label{subsec:modeling}

In order to quantify the CSM properties of \sn{}, we compared the spectral and photometric properties of \sn{} to a model grid of non-LTE, radiative transfer simulations covering a wide range of progenitor mass-loss rates ($\dot{M} = 10^{-6} - 10^{0}~\Msun$~yr$^{-1}$; $v_w = 50~\kms$) and CSM radii ($R = 10^{14} - 10^{16}$~cm), all in spherical symmetry. Simulations of the SN ejecta-CSM interaction were performed with the multi-group radiation-hydrodynamics code \heracles\ \citep{gonzalez_heracles_07,vaytet_mg_11,D15_2n}, which consistently computes the radiation field and hydrodynamics. Then, at selected snapshots in time post-explosion, the hydrodynamical variables are imported into the non-LTE radiative-transfer code \cmfgen\ \citep{hillier12, D15_2n} for an accurate calculation of the radiative transfer, which includes a complete model atom, $\sim10^6$ frequency points, and treatment of continuum and line processes as well as electron scattering. For each model, we adopt an explosion energy of $1.2\times 10^{51}$~erg, a 15$\Msun$ progenitor with a radius ranging from $R_{\star} \approx 500 - 700~\Rsun$, and a CSM composition set to the surface mixture of a RSG progenitor \citep{davies_dessart_19}. For the simulations presented in this work, the CSM extent is much greater than $R_{\star}$ ($\sim$500--1200~$\Rsun$ for a RSG mass range of $\sim$10--20~$\Msun$) and therefore the progenitor properties have no impact during phases of ejecta-CSM interaction. The progenitor radius plays a more significant role on the light curve evolution during the plateau phase (e.g., see \citealt{d13_sn2p,wjg22}), i.e., once the interaction phase is over and the emission from from the deeper ejecta layers dominate the SN luminosity. Specific methods for each simulation can be found in \cite{Dessart16, dessart17, wjg22, Dessart23} and all CSM properties of each model are presented in Table \ref{tab:models}. 



\begin{figure}[t!]
\centering
\includegraphics[width=0.47\textwidth]{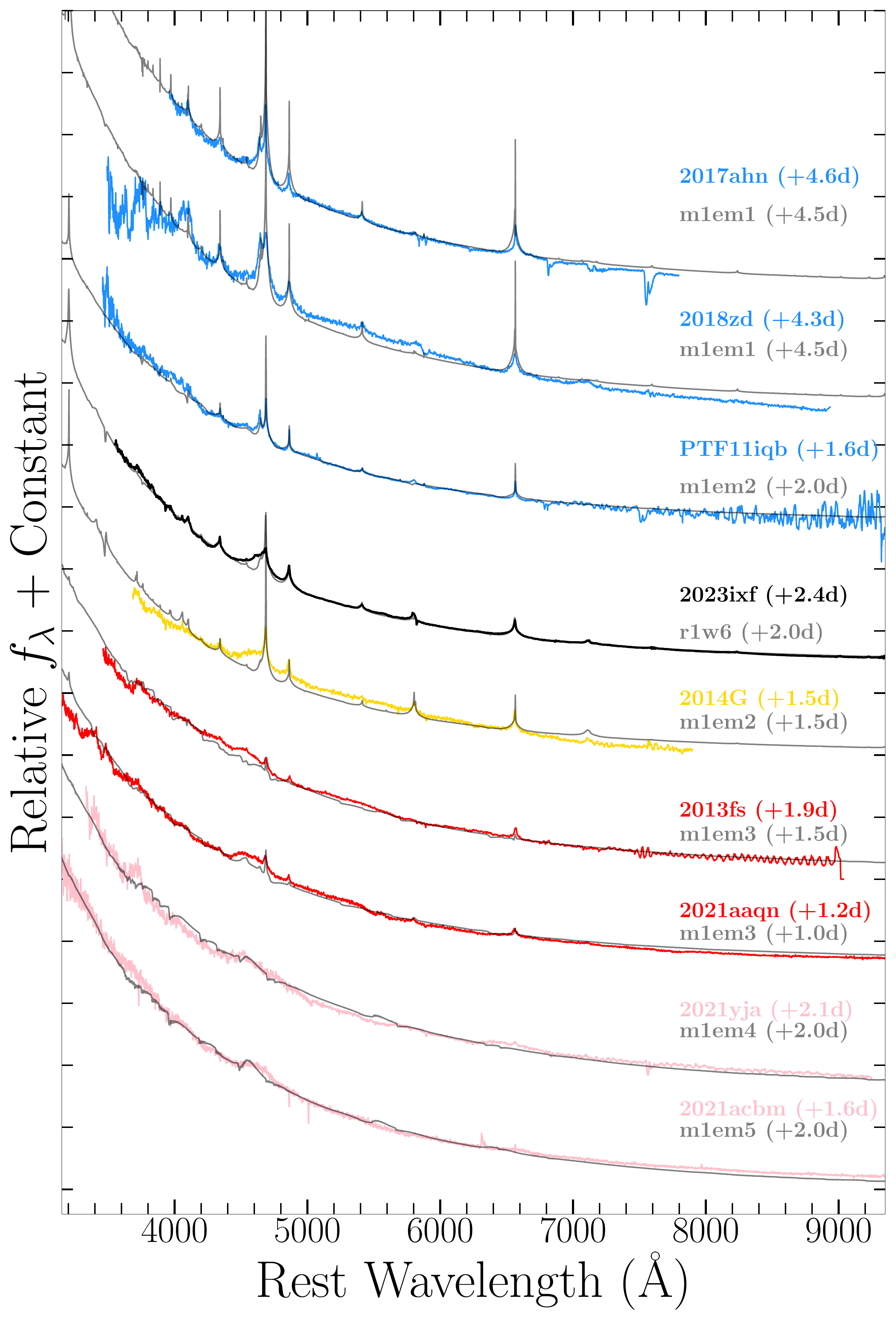}
\caption{Early-time spectra of SNe~II with varying degrees of CSM interaction, selected from \cite{Jacobson-Galan23}. Colors delineate differences between duration of IIn-like signatures as well as the presence of different high-ionization species at phases of $t \approx 2$~days (e.g., spectra in blue show \ion{N}{III} while others do not). Overplotted in grey are best fitting \cmfgen\ models for high ($10^{-1}~\Msun$~yr$^{-1}$; top) to low ($10^{-5}~\Msun$~yr$^{-1}$; bottom) mass-loss rates. \sn{} lies within the continuum of CSM-interacting SNe~II and the evolution of its prominent spectral signatures of photo-ionized CSM is consistent with a progenitor mass-loss rate of $\dot{M} = 10^{-2}~\Msun$~yr$^{-1}$. \label{fig:spec_compare}}
\end{figure}

\begin{figure*}
\centering
\subfigure[]{\includegraphics[width=0.32\textwidth]{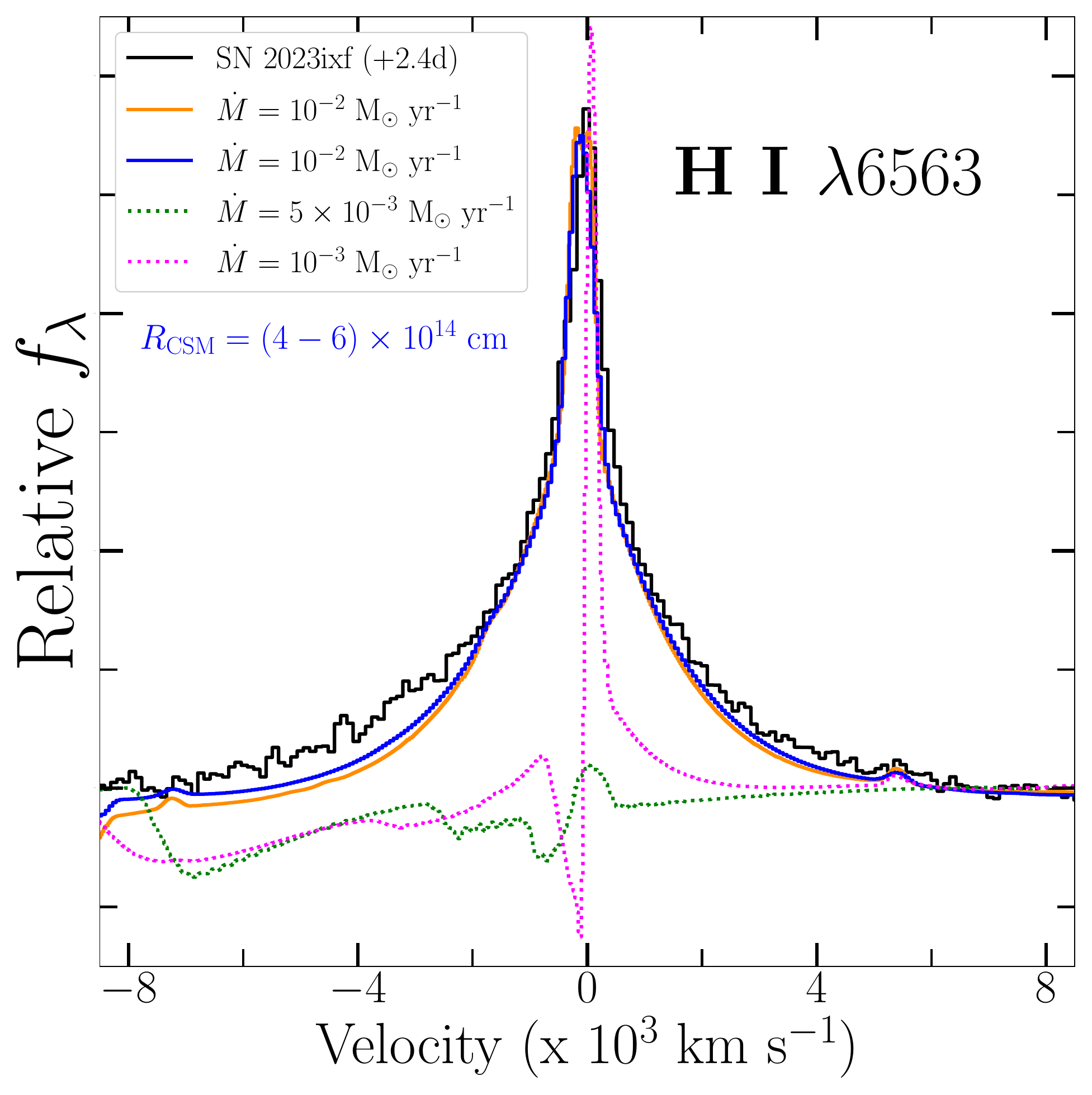}}
\subfigure[]{\includegraphics[width=0.32\textwidth]{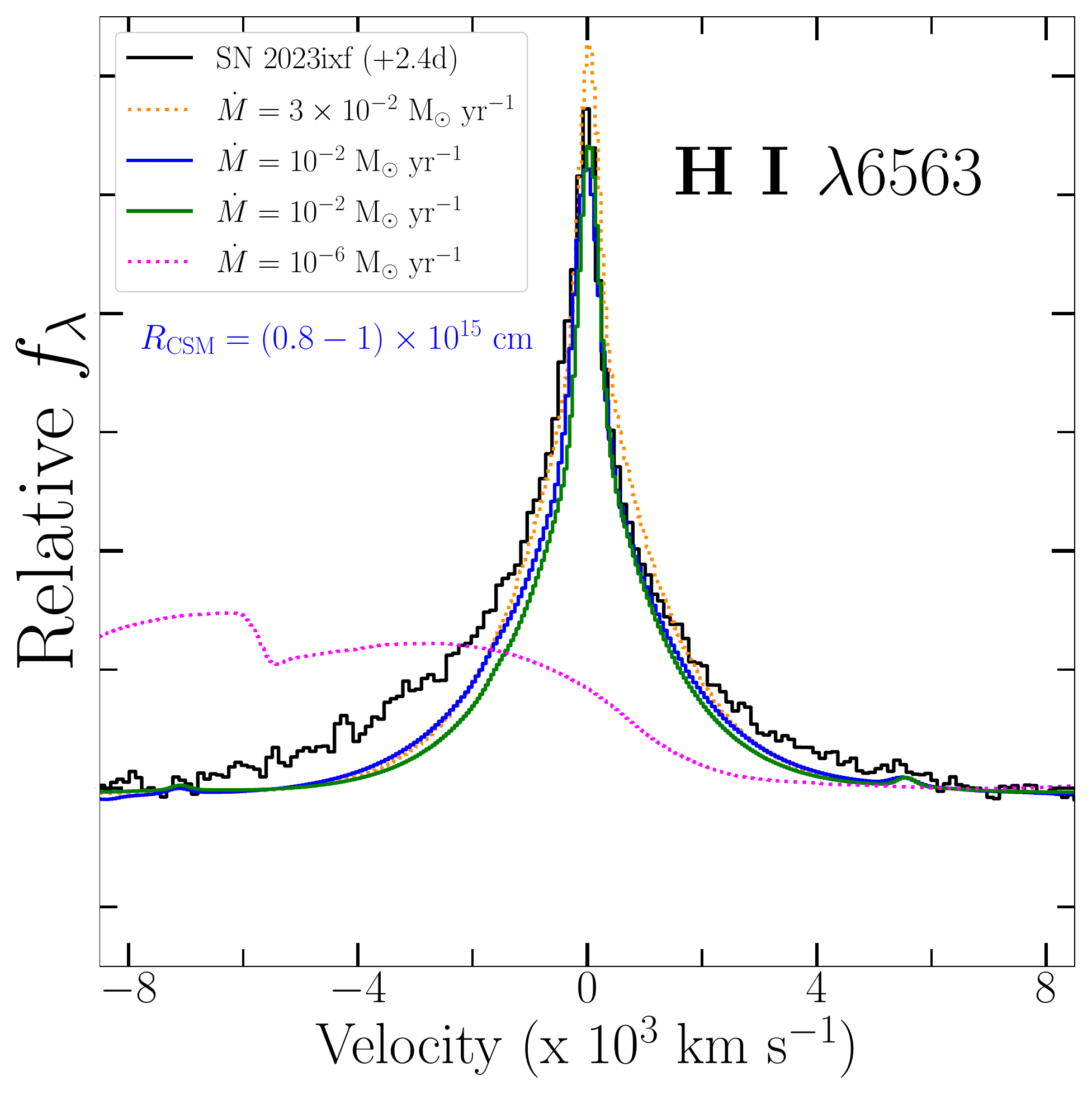}}
\subfigure[]{\includegraphics[width=0.32\textwidth]{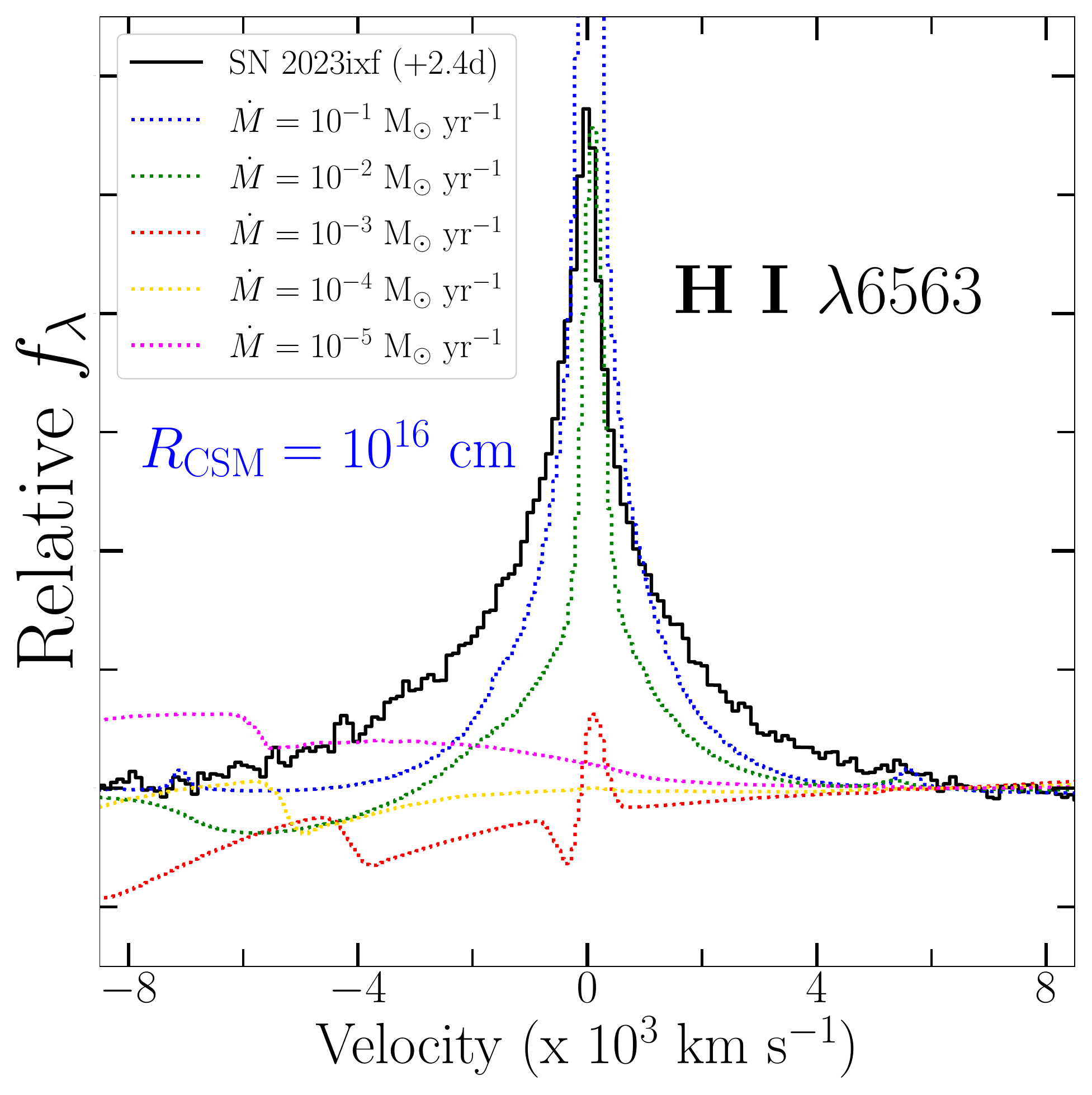}}\\
\subfigure[]{\includegraphics[width=0.32\textwidth]{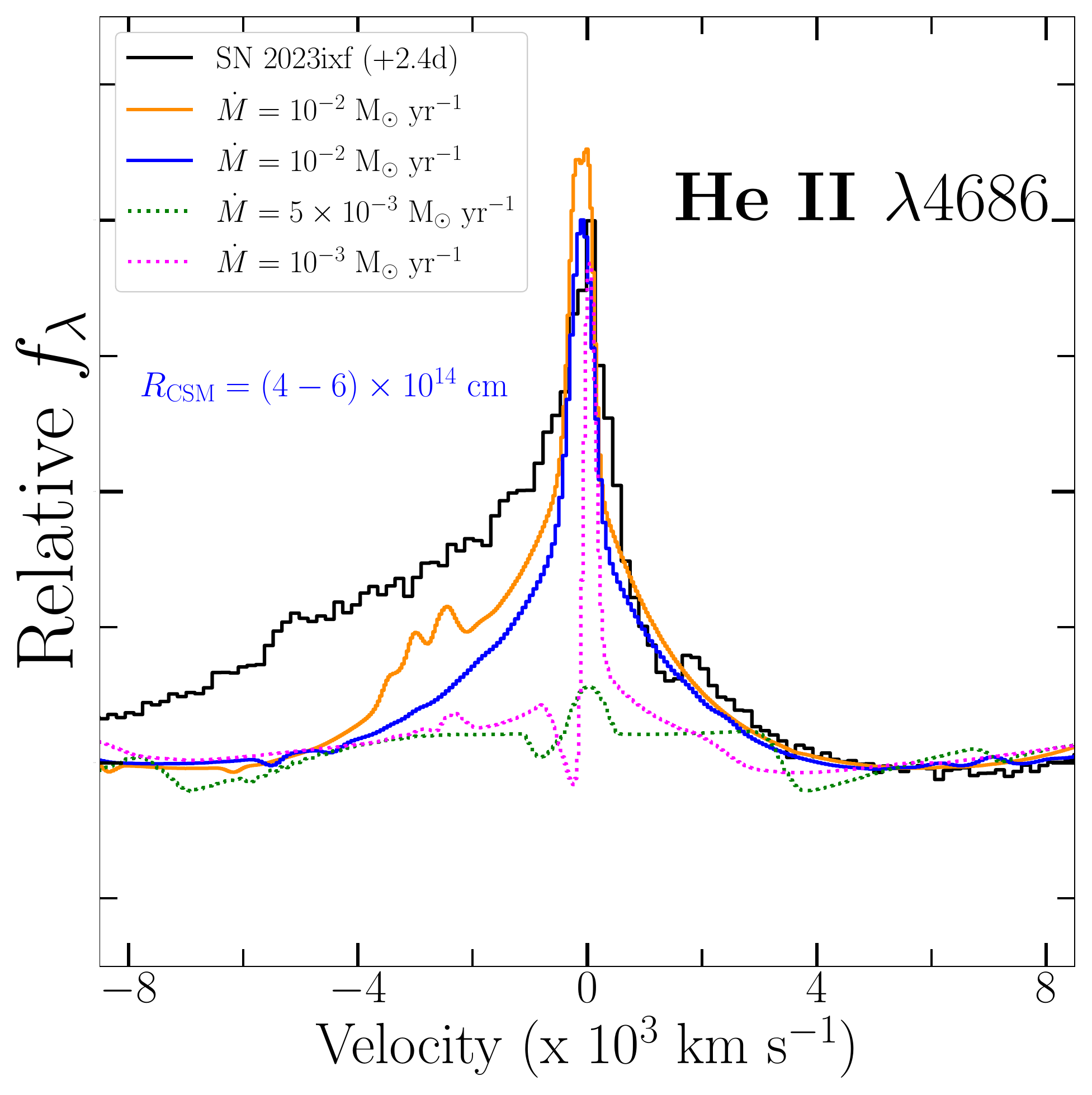}}
\subfigure[]{\includegraphics[width=0.32\textwidth]{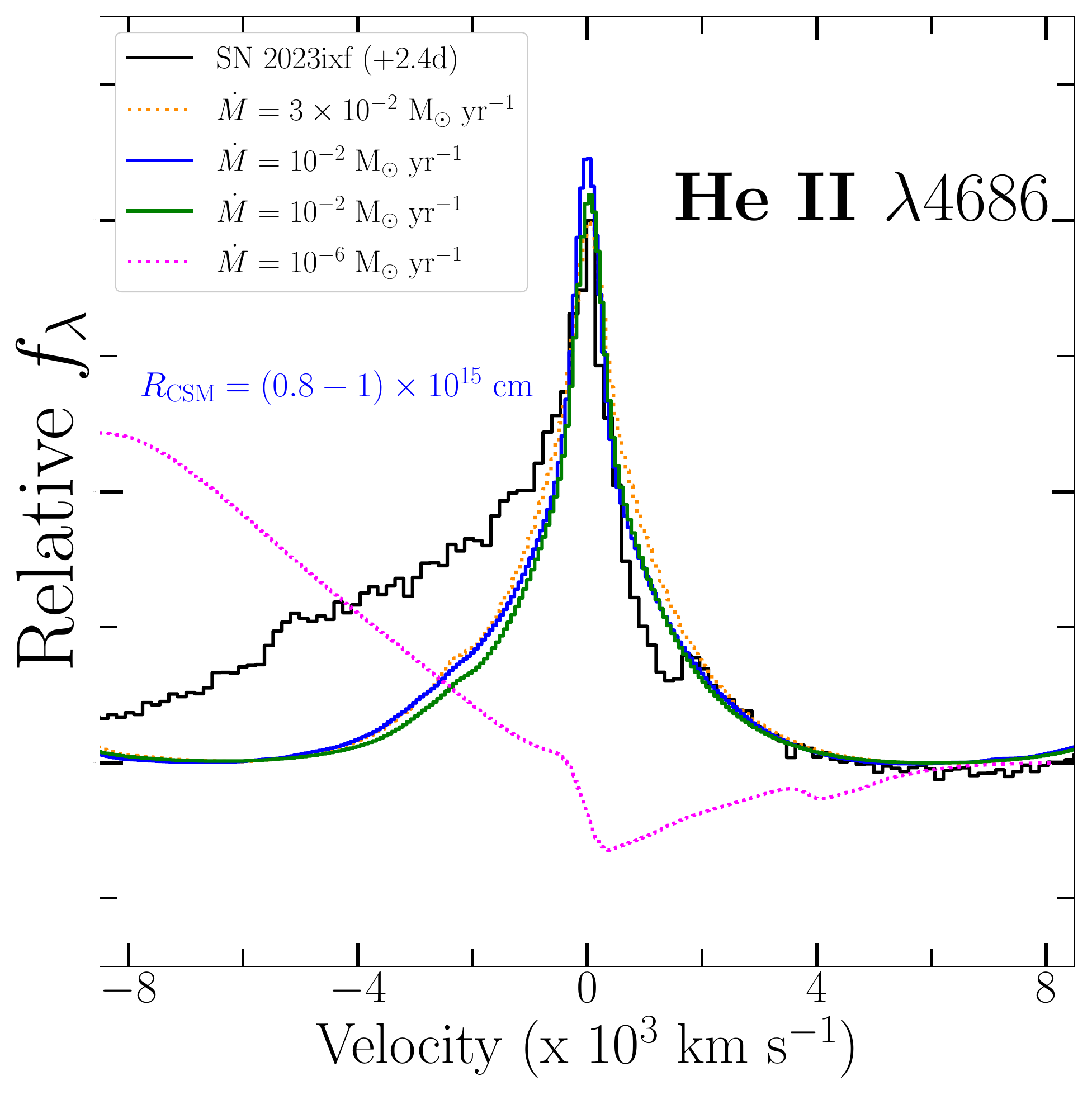}}
\subfigure[]{\includegraphics[width=0.32\textwidth]{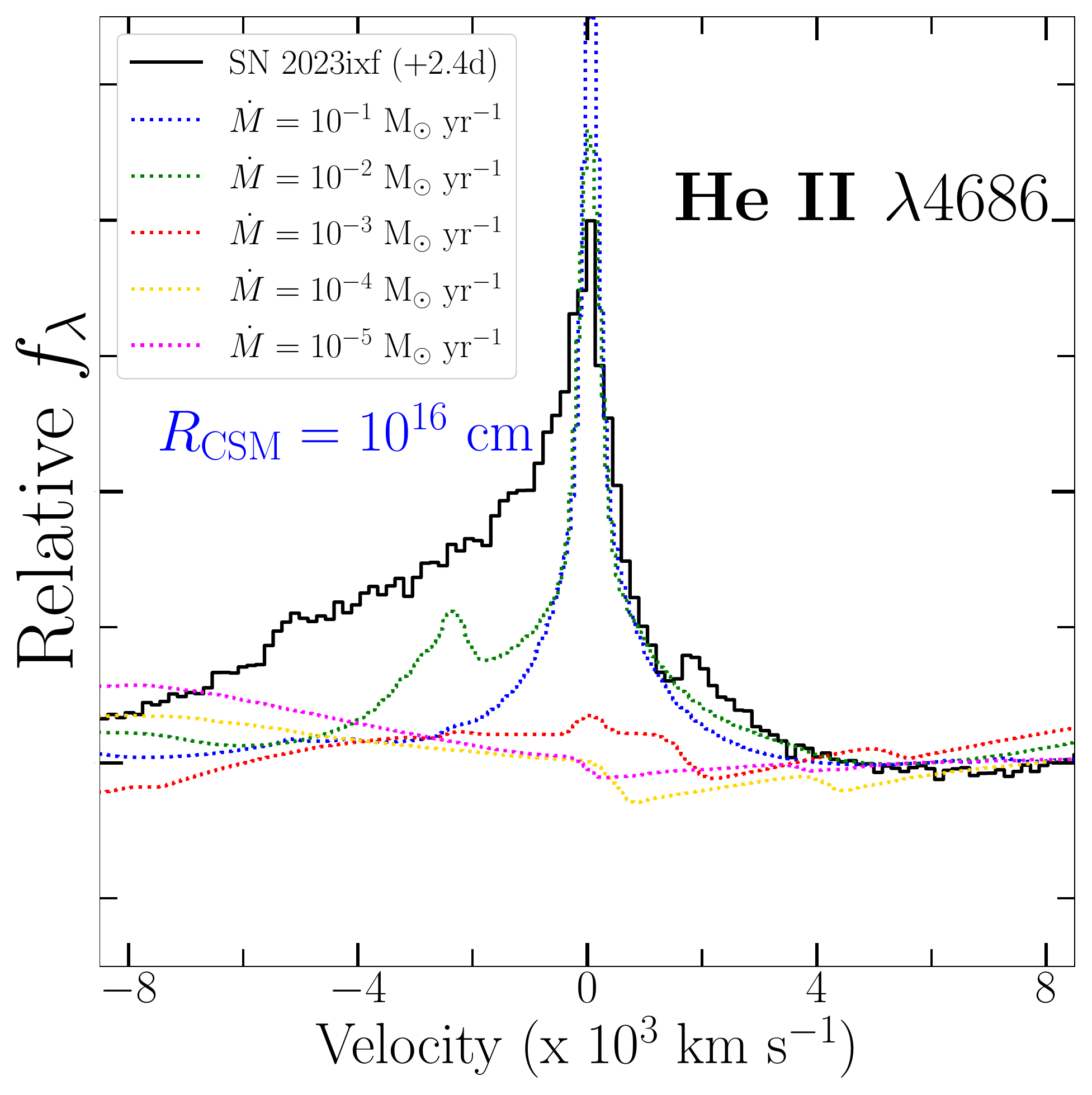}}
\caption{H$\alpha$ velocity of \sn{} at $\delta t = 2.4$~days (black) with respect to \cmfgen\ model spectra at +2~days post-explosion. Best fit models (e.g., \S\ref{subsec:modeling}) shown as solid lines; additional, inconsistent models plotted as dotted lines. For proper comparison, models are scaled to the continuum of \sn{} and the peak of the presented emission line profile. These models have varying mass-loss rates that are contained within CSM radii of (a) $R_{\rm CSM} = (4-6) \times 10^{14}$~cm, (b) $R_{\rm CSM} = (0.8-1) \times 10^{15}$~cm, and (c) $R_{\rm CSM} = 10^{16}$~cm. (d)/(e)/(f) Same as above, but for \ion{He}{ii} velocities.  
\label{fig:model_all} }
\end{figure*}

From the \cmfgen\ model grid, we identify four models (r1w6, r1w6a,b,c) with the smallest residuals between model predictions and both the observed multi-color peak magnitudes and rise-times (\S\ref{subsec:phot_properties}) as well as the duration of IIn profiles (\S\ref{subsec:spec_analysis}). The best matched model peak magnitudes are within $\sim$0.5~mag of \sn{} in all filters and have a $t_{\rm IIn}$ that is within $\pm 3$~days of that observed in \sn{}. The features used for determination of the most consistent model are presented in Figure \ref{fig:max_lines}. We find that the best-fit models to \sn{} have a mass-loss rate of $\dot{M} = 10^{-2}~\Msun$~yr$^{-1}$, confined to a radius of $r = (0.5-1)\times 10^{15}$~cm and containing a total CSM mass of $M_{\rm CSM} \approx 0.04 - 0.07~\Msun$. Based on model predictions, the mass-loss then decreases to $\dot{M} = 10^{-6}~\Msun$~yr$^{-1}$ (e.g., Fig. \ref{fig:LC_rho}b) at larger distances ($r > 10^{15}$~cm) with a constant wind velocity of $v_w = 50~\kms$; this geometry is consistent with the changing X-ray absorption observed in \sn{} \citep{Grefenstette23}. This wind velocity is not derived from spectroscopic observations but the narrow line cores of the higher resolution Kast spectra indicate $v_w \lesssim 150~\kms$. The spectral time series of the r1w6b model and multi-color light curve of the r1w6a,b,c models are presented in Figures \ref{fig:spec_series} and \ref{fig:LC_rho}(a). In Figure \ref{fig:model_all}, we present \sn{} spectra with respect to a range of early-time \cmfgen\ models with varying mass-loss rates and CSM radii to further illustrate the consistencies and inconsistencies between models and observations. In Figure \ref{fig:spec_UV}, we present the UV spectrum of the r1w6b at +2.5~days, which predicts a plethora of high-ionization features (e.g., \ion{O}{iv/vi}) in the near-to-far UV spectra of \sn{}. Furthermore, the r1w6b model adequately reproduces the observed IIn-like emission line profiles in \sn{} (e.g., Figs. \ref{fig:spec_compare} \& \ref{fig:model_all}) using the $15~\Msun$, solar metallicity, RSG progenitor model composition \citep{dessart17, davies_dessart_19}.


\begin{table}[t!]
\begin{center}
\caption{Main parameters of \sn{} \label{tbl:params}}
\vskip0.1in
\begin{tabular}{lccc}
\hline
\hline
Host Galaxy &  &  M101 \\ 
Redshift &  &  0.000804\\  
Distance &  &  6.9~Mpc\footnote{\cite{riess22}}\\ 
Time of First Light (MJD) &  &  60082.83\\
$E(B-V)_{\textrm{MW}}$ &  &  0.008~mag\footnote{\cite{schlegel98,schlafly11}}\\
$E(B-V)_{\textrm{host}}$ &  &  0.033~mag\footnote{\cite{Poznanski12}}\\
$M_{u}^{\mathrm{peak}}$[$t_r$] &  &  -18.6~mag[4.9~d]\\
$M_{B}^{\mathrm{peak}}$[$t_r$] &  &  -18.5~mag[5.7~d]\\
$M_{g}^{\mathrm{peak}}$[$t_r$] &  &  -18.4~mag[5.8~d]\\
$M_{V}^{\mathrm{peak}}$[$t_r$] &  &  -18.1~mag[6.0~d]\\
$M_{r}^{\mathrm{peak}}$[$t_r$] &  &  -18.0~mag[6.1~d]\\
$M_{i}^{\mathrm{peak}}$[$t_r$] &  &  -17.9~mag[7.8~d]\\
$M_{z}^{\mathrm{peak}}$[$t_r$] &  &  -17.8~mag[8.2~d]\\
R$_{\rm CSM}$ &  &  $(0.5-1) \times 10^{15}$~cm\\
M$_{\rm CSM}$ &  &  $(0.04-0.07)~\Msun$\\
$\dot{M}$[$v_w$]\footnote{Mass-loss within $r<10^{15}$~cm} &  &  $10^{-2}~\Msun$~yr$^{-1}$[$50~\kms$]\\
CSM Composition &  &  Solar Metallicity\footnote{Not varied in model grid}\\
Time of $\dot{M}$ &  &  $\sim$3-6~years pre-SN\\
\hline
\end{tabular}
\end{center}
\label{table:props}
\end{table}

Unlike the other CSM structures explored in the \cmfgen\ model grid, the r1w6b model best reproduces both the observed peak absolute magnitudes in $uBgVriz$ filters as well as the duration of IIn-like line profiles observed in \sn{} (Figure \ref{fig:max_lines}). However, this model cannot match the early light curve slope observed in \sn{}, which is likely a result of the density profile invoked. A better fit to the light curve would require a higher density immediately above $R_{\star}$ (e.g., through a larger scale height) and a more gradual decline in density at the outer edge of the dense CSM i.e., at $\sim 8\times10^{15}$~cm. A more extended CSM, as in model r1w6c, increases the rise time to maximum and is thus not a suitable solution. Furthermore, a larger model kinetic energy will also increase the luminosity at early-times, which would provide more consistency between the model light curve rise and \sn{}. 

The evolution of the line profiles in the r1w6b model are consistent with the observed transition from electron-scattering broadened line profiles of Balmer series \ion{H}{i}, \ion{He}{ii}, \ion{C}{iv}, and \ion{N}{iii/iv} ($t<7$~days) to Doppler broadened absorption profiles of the fastest moving H-rich SN ejecta ($t>7$~days). Nevertheless, for a consistent continuum slope, the model spectra over-predict the narrow line emission observed in \sn{} (e.g., Fig. \ref{fig:spec_series}), which is likely caused by a smaller emitting volume of dense CSM in \sn{} than the r1w6b model. The line strengths of \sn{} are well-matched by the r1w6 model, which is characterized by the same $\dot M$ but smaller CSM radius (e.g., Fig. \ref{fig:spec_compare}); this model, however, is unable to reproduce the extended duration of the IIn-like features (e.g., see Fig. \ref{fig:max_lines}). Furthermore, as shown in Figure \ref{fig:vels}(c), the \ion{He}{ii}~$\lambda4686$ line profile is not completely reproduced by the model spectrum, which could be due to the fact that these simulations are performed in 1D, assume spherical symmetry of the CSM, or require higher kinetic energies. Additionally, once the IIn-like profiles fade, the model \ion{H}{i} ejecta velocities are lower than in \sn{} (e.g., Fig. \ref{fig:vels}). It should also be noted that the narrow H$\alpha$ P Cygni profile that develops in SN~2023ixf at $t>8$~days has higher velocities ($\sim 100~\kms$) than the r1w6b model ($\sim50~\kms$), suggesting potential radiative acceleration of the unshocked CSM.

Using the lower limit on the SN shock velocity of $\gtrsim$8500~$\kms$ as observed at the bluest edge of the H$\alpha$ absorption profile shown in Figure \ref{fig:vels}, we find that the location of the SN shock at $t = 6.4$~days is $r \gtrsim 5\times 10^{14}$~cm, which corresponds to an optical depth to electron-scattering of $\tau \approx 10$ in the r1w6b model. Then, by $t = 10.6$~days, the shock is located at $r \gtrsim 8\times 10^{14}$~cm, which in the r1w6b model, is in CSM with an electron-scattering optical depth of $\tau \approx 0.2$. This decrease in $\tau_{\rm ES}$ in the r1w6b model is consistent with the observed fading of IIn-like profiles in \sn{} and reflects a reduction in CSM density at $r > 10^{15}$~cm. However, it should be noted that the shock velocity decreases as the shock crosses the CSM and, therefore, the shock position is not as simplified as $R_{\rm sh} = v_{\rm sh} \times t$. We present the time-series evolution of the r1w6b model luminosity, density, temperature and velocity as a function of radius in Figure \ref{fig:model_props}. Furthermore, at +15~days, the r1w6b model has a maximum ejecta velocity of $\sim 6500~\kms$, which is below the maximum velocities observed in the H$\alpha$ profile of \sn{}. This may indicate some degree of CSM asymmetry that would cause a deceleration along certain lines of sight while also allowing for typical SN ejecta velocities of $\sim 10^4~\kms$ to be preserved. Also, the weaker Doppler broadened absorption observed in \sn{} compared to the r1w6b model may be the result of persistent CSM interaction that will contribute weak, broad, and boxy H$\alpha$ emission capable of reducing the absorption profile strength \citep{dessart22}.

\section{Discussion} \label{sec:discussion}

The first 2 weeks of photometric and spectroscopic observations of \sn{} have revealed essential characteristics of the SN progenitor system and the explosion itself. Based on the best fitting \cmfgen\ model, the progenitor of \sn{} was likely a RSG with a mass-loss rate of $\dot{M} \approx 10^{-2}~\Msun$~yr$^{-1}$, which created dense CSM extending to $r \approx (0.5 - 1)\times10^{15}$~cm that contained a total mass of $M_{\rm CSM} \approx 0.04 - 0.07~\Msun$. Furthermore, we find that the observed light curve can be fit with standard explosion energy ($1.2\times 10^{51}$~ergs) and that the IIn-like signatures in \sn{} can be modeled with a CSM composition that matches typical RSG surface abundances with no need for significant N or He enrichment.

For a wind velocity of $\sim 50~\kms$, the proposed CSM extent translates to a period of enhanced mass-loss (i.e. ``super wind'') in the last $\sim$3-6~years years prior to core collapse. This scenario comports with the observed duration of \sn{}'s IIn-like line profiles ($\sim$8~days), after which time the optical depth to electron-scattering in the CSM has decreased and \sn{} begins to show absorption profiles from the outer, H-rich ejecta and fast moving dense shell. If the CSM detected in \sn{} represents the only high-density shell of CSM (i.e. only one super wind phase), then the SN shock should continue to sample low-density material ($\dot{M} \approx 10^{-6}~\Msun$~yr$^{-1}$, $v_w = 50~\kms$) at larger distances ($r>10^{15}$~cm). Overall, both the confined high-density CSM shell and the extended low-density wind may have made the RSG progenitor star quite dust obscured prior to explosion \citep{davies22}. This is consistent with the findings of \cite{Kilpatrick23} who show that the pre-explosion {\it Hubble} and {\it Spitzer} imaging of \sn{} indicates a moderately sized ($\sim$11~$\Msun$) RSG progenitor star enshrouded in a dust shell. 

This physical progenitor picture is also consistent with the initial findings of multi-wavelength observations of \sn{}. X-ray observations revealed a 3-30\,keV spectrum consistent with bremsstrahlung emission that initially showed significant absorption in the soft part of the spectrum \citep{Grefenstette23}. These observations are indicative of shock interaction with dense CSM in a confined shell. Furthermore, radio observations have so far produced non-detections at $\nu = 1-230$~GHz with SMA, GMRT, and VLA \citep{Berger23, Matthews23, chandra23_radio}, likely caused by large free-free absorption in the optically thick CSM. However, as the SN shock now enters lower density material at larger distances, it is probable that the radio emission in \sn{} will become detectable.

\section{Conclusions} \label{sec:conclusion}

In this paper we have presented UV/optical observations and models of the nearby SN~II, 2023ixf located in nearby spiral host galaxy M101 at $d=6.9$~Mpc. Below we summarize the primary observational findings of \sn{}:  

\begin{itemize}

\item The early-time spectra of \sn{} show prominent narrow emission lines of \ion{H}{i}, \ion{He}{i/ii}, \ion{N}{iii/iv/v} and \ion{C}{iv} that result from the photo-ionization of dense, optically thick CSM. These electron-scattering broadened profiles (i.e., IIn-like) last for $t_{\rm IIn} \approx 8$~days, after which time broad absorption profiles from the fastest H-rich SN ejecta begin to form.


\item CSM interaction in \sn{} caused increased peak absolute magnitudes (e.g., $M_u = -18.6$~mag, $M_g = -18.4$~mag) relative to SNe~II that occur in a low density circumstellar environment (e.g., $\rho < 10^{-16}$~g~cm$^{-3}$). The observed multi-color peak absolute magnitudes and duration of the IIn-like profiles (i.e., $t_{\rm IIn})$ places \sn{} in the continuum of SNe~II with varying degrees of RSG mass-loss before explosion. Compared to the sample of CSM-interacting SNe~II compiled in \citet{Jacobson-Galan23}, \sn{} is most similar to SNe~2017ahn, 2018zd, 2020pni and 2022ffg. 

\item Comparing \sn{}'s peak absolute magnitudes and duration of IIn-like profiles to a grid of \cmfgen\ simulations suggests a CSM that has a composition typical of a solar-metallicity RSG, is confined to $r \approx (0.5-1)\times 10^{15}$~cm, and is formed from wind corresponding to a progenitor mass-loss rate of $\dot{M} = 10^{-2}~\Msun$~yr$^{-1}$ (i.e., $\rho \approx 10^{-12}$g~cm$^{-3}$ at $r = 10^{14}$~cm). Adopting a wind velocity of $v_w = 50~\kms$, this scenario corresponds to a period of enhanced mass-loss (i.e., ``superwind'') during the last $\sim$3-6~years years before core-collapse. 

\end{itemize}

Given its close proximity and present brightness, \sn{} is poised to become the best studied CSM-interacting SN~II to date. Future, multi-wavelength observations will, among other things, uncover the density profile of the confined CSM as well as the mass-loss history of the RSG progenitor in its final decades to centuries. 

\section{Acknowledgements} \label{Sec:ack}

Research at UC Berkeley is conducted on the territory of Huichin, the ancestral and unceded land of the Chochenyo speaking Ohlone people, the successors of the sovereign Verona Band of Alameda County. PS1/2 observations were conducted on the stolen land of the k\={a}naka `\={o}iwi people. We stand in solidarity with the Pu'uhonua o Pu'uhuluhulu Maunakea in their effort to preserve these sacred spaces for native Hawai`ians. Shane 3-m observations were conducted on the stolen land of the Ohlone (Costanoans), Tamyen and Muwekma Ohlone tribes.

We thank Greg Zeimann for help with HET data reduction and Dan Weisz for Kast observations. 

The Young Supernova Experiment and its research infrastructure is supported by the European Research Council under the European Union's Horizon 2020 research and innovation programme (ERC Grant Agreement No.\ 101002652, PI K.\ Mandel), the Heising-Simons Foundation (2018-0913, PI R.\ Foley; 2018-0911, PI R.\ Margutti), NASA (NNG17PX03C, PI R.\ Foley), NSF (AST-1720756, AST-1815935, PI R.\ Foley; AST-1909796, AST-1944985, PI R.\ Margutti), the David \& Lucille Packard Foundation (PI R.\ Foley), VILLUM FONDEN (project number 16599, PI J.\ Hjorth), and the Center for AstroPhysical Surveys (CAPS) at the National Center for Supercomputing Applications (NCSA) and the University of Illinois Urbana-Champaign.

W.J.-G.\ is supported by the National Science Foundation Graduate Research Fellowship Program under Grant No.~DGE-1842165. W.J.-G.\ acknowledges support through NASA grants in support of {\it Hubble Space Telescope} program GO-16075 and 16500. This research was supported in part by the National Science Foundation under Grant No. NSF PHY-1748958.  R.M.\ acknowledges support by the National Science Foundation under Award No. AST-2221789
and AST-2224255.  The Margutti team at UC Berkeley is partially funded by the Heising-Simons Foundation under grant \# 2018-0911 and \#2021-3248 (PI: Margutti).

C.D.K.\ is partly supported by a CIERA postdoctoral fellowship and a grant in support of NASA program HST-GO-16136. V.A.V.\ acknowledges support by the National Science Foundation under Award No.AST-2108676. C.R.A.\ was supported by grants from VILLUM FONDEN (project numbers 16599). Parts of this research were supported by the Australian Research Council Centre of Excellence for All Sky Astrophysics in 3 Dimensions (ASTRO 3D), through project number CE170100013. Y.-C.P.\ is supported by the National Science and Technology Council (NSTC grant 109-2112-M-008-031-MY3. A.G.\ is supported by the National Science Foundation Graduate Research Fellowship Program under Grant No. DGE–1746047. A.G.\ also acknowledges funding from the Center for Astrophysical Surveys Fellowship at UIUC/NCSA and the Illinois Distinguished Fellowship. C.G. is supported by a VILLUM FONDEN Young Investigator Grant (project number 25501). SJS acknowledges funding from STFC grants Ref: ST/T000198/1 and ST/S006109/1. J.H. was supported by a VILLUM FONDEN Investigator grant (project number 16599). D.~M.\ acknowledges NSF support from grants PHY-2209451 and AST-2206532.

The UCSC team is supported in part by NASA grants NNG17PX03C and 80NSSC22K1518, NSF grant AST--1815935, and by a fellowship from the David and Lucile Packard Foundation to R.J.F.

YSE-PZ \citep{Coulter23} was developed by the UC Santa Cruz Transients Team with support from NASA grants NNG17PX03C, 80NSSC19K1386, and 80NSSC20K0953; NSF grants AST-1518052, AST-1815935, and AST-1911206; the Gordon \& Betty Moore Foundation; the Heising-Simons Foundation; a fellowship from the David and Lucile Packard Foundation to R.J.F.; Gordon and Betty Moore Foundation postdoctoral fellowships and a NASA Einstein fellowship, as administered through the NASA Hubble Fellowship program and grant HST-HF2-51462.001, to D.O.J.; and a National Science Foundation Graduate Research Fellowship, administered through grant No.\ DGE-1339067, to D.A.C.

A major upgrade of the Kast spectrograph on the Shane 3~m telescope at Lick Observatory, led by Brad Holden, was made possible through generous gifts from the Heising-Simons Foundation, William and Marina Kast, and the University of California Observatories. Research at Lick Observatory is partially supported by a generous gift from Google.

This research was supported by the Munich Institute for Astro-, Particle and BioPhysics (MIAPbP) which is funded by the Deutsche Forschungsgemeinschaft (DFG, German Research Foundation) under Germany´s Excellence Strategy – EXC-2094 – 390783311.

The Pan-STARRS1 Surveys (PS1) and the PS1 public science archive have been made possible through contributions by the Institute for Astronomy, the University of Hawaii, the Pan-STARRS Project Office, the Max-Planck Society and its participating institutes, the Max Planck Institute for Astronomy, Heidelberg and the Max Planck Institute for Extraterrestrial Physics, Garching, The Johns Hopkins University, Durham University, the University of Edinburgh, the Queen's University Belfast, the Harvard-Smithsonian Center for Astrophysics, the Las Cumbres Observatory Global Telescope Network Incorporated, the National Central University of Taiwan, STScI, NASA under grant NNX08AR22G issued through the Planetary Science Division of the NASA Science Mission Directorate, NSF grant AST--1238877, the University of Maryland, Eotvos Lorand University (ELTE), the Los Alamos National Laboratory, and the Gordon and Betty Moore Foundation.

IRAF is distributed by NOAO, which is operated by AURA, Inc., under cooperative agreement with the National Science Foundation (NSF).

\facilities{YSE/PS1/PS2, Shane Telescope (Kast), Nickel Telescope, Thailand Robotic Telescope, Auburn Telescope, Las Cumbres Observatory, Lulin Telescope, Hobby Eberly Telescope (LRS), Apache Point Observatory (KOSMOS)}

\software{IRAF (Tody 1986, Tody 1993),  photpipe \citep{Rest+05}, DoPhot \citep{Schechter+93}, HOTPANTS \citep{becker15}, YSE-PZ \citep{Coulter22, Coulter23}, \cmfgen\ \citep{hillier12, D15_2n}, \heracles\ \citep{gonzalez_heracles_07,vaytet_mg_11,D15_2n} }

\bibliographystyle{aasjournal} 
\bibliography{references} 


\clearpage
\appendix

\begin{figure*}
\centering
\includegraphics[width=0.8\textwidth]{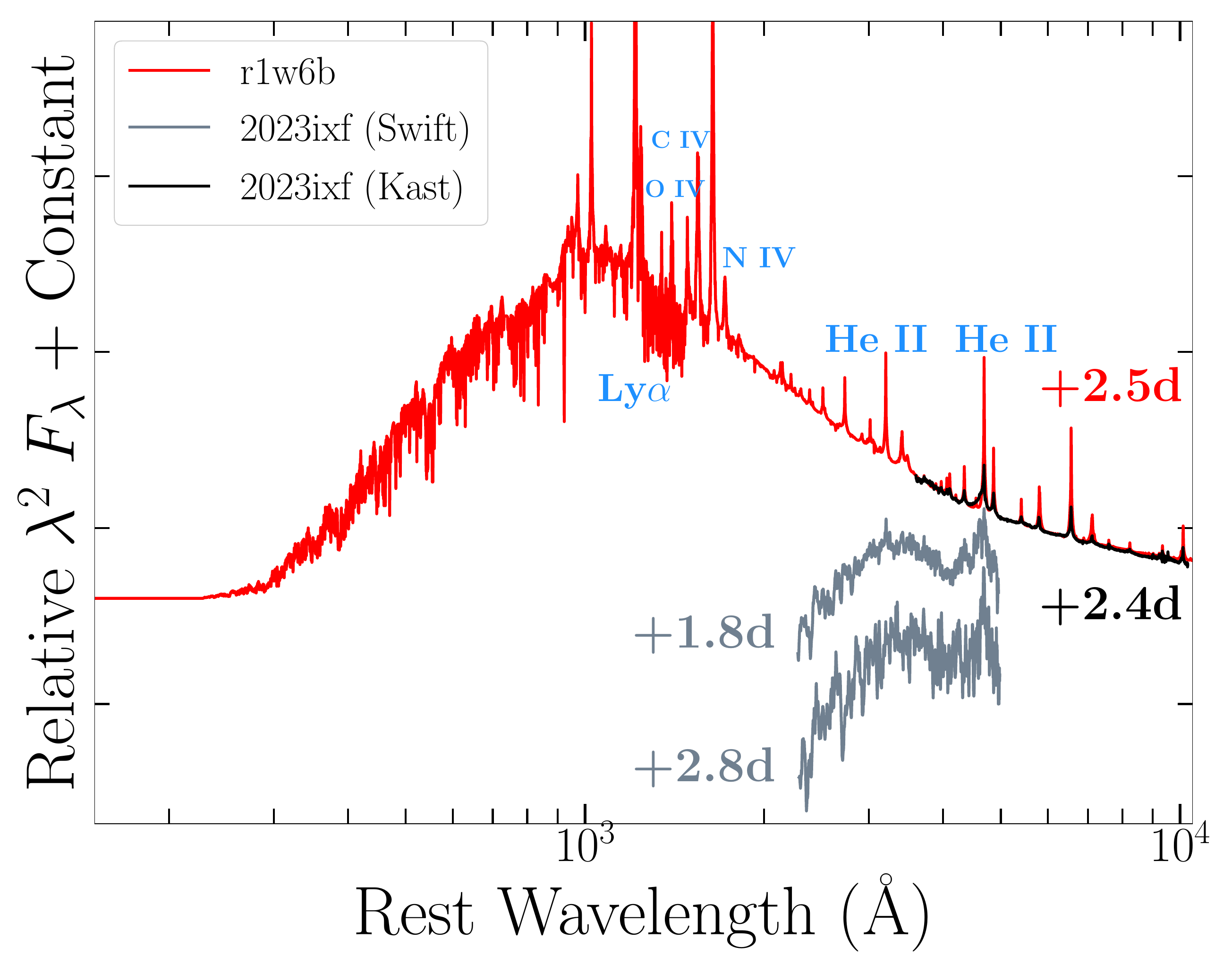}
\caption{{\it Swift} UV grism spectra (grey) with respect to best fitting \cmfgen\ model (red); y-axis is in units of $\lambda^2~F_{\lambda}$. The {\it Swift} UV grism response function below $\sim3200$~\AA~is not reflective of the true slope of the SN SED. The r1w6b model shows that the NUV and FUV spectra of \sn{} likely contains a plethora of narrow, high-ionization emission lines (e.g., \ion{He}{ii}, \ion{C}{iv}, \ion{N}{iv/v}, and \ion{O}{iv}) derived from CSM interaction.  \label{fig:spec_UV}}
\end{figure*}

\begin{figure*}
\centering
\subfigure[]{\includegraphics[width=0.49\textwidth]{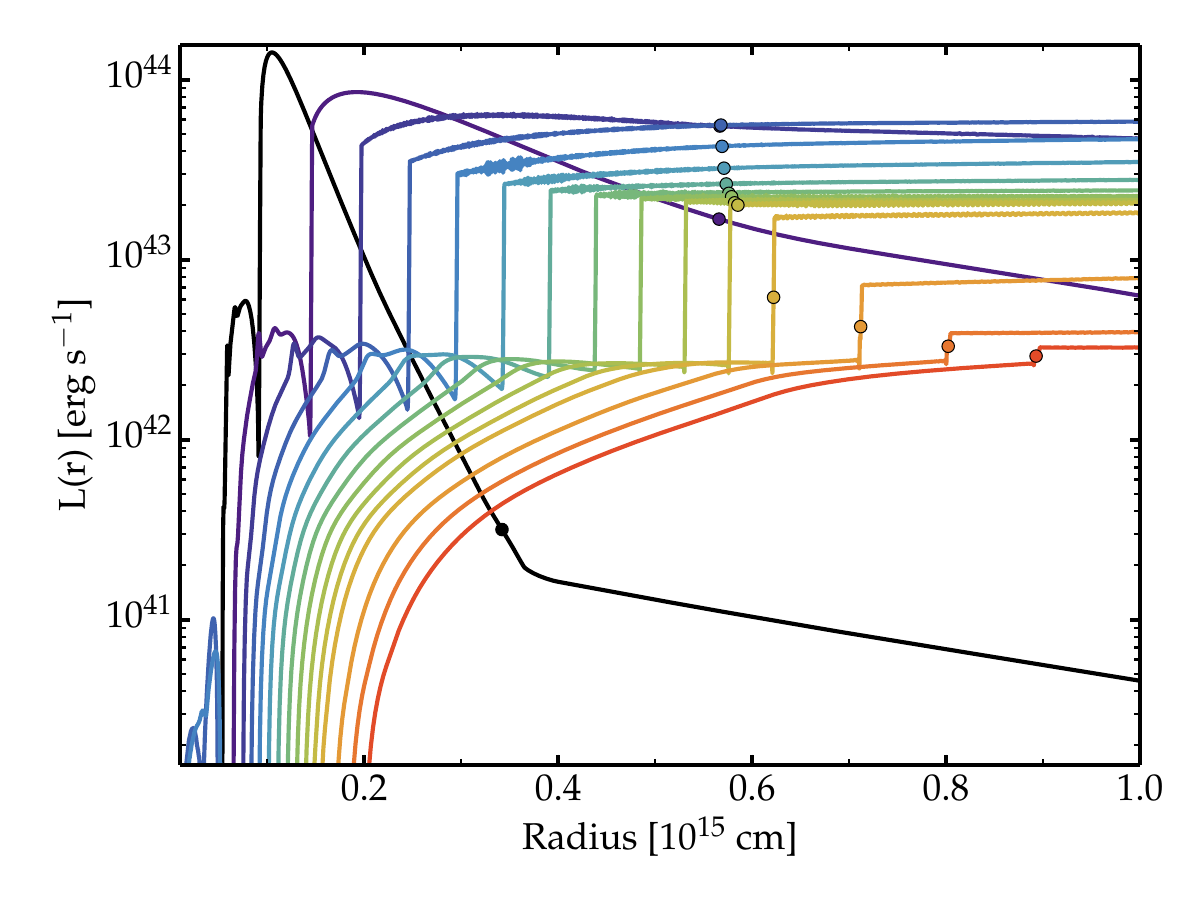}}
\subfigure[]{\includegraphics[width=0.49\textwidth]{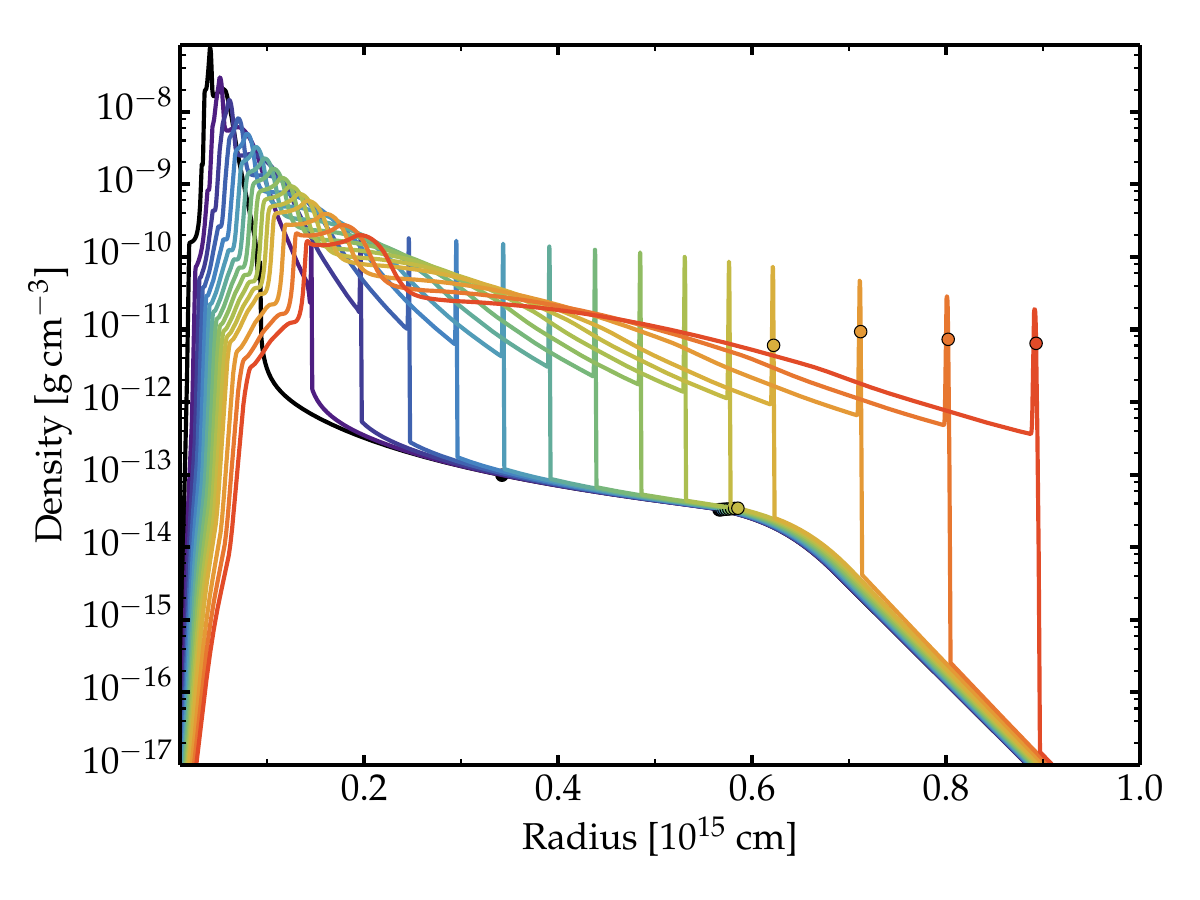}}\\
\subfigure[]{\includegraphics[width=0.49\textwidth]{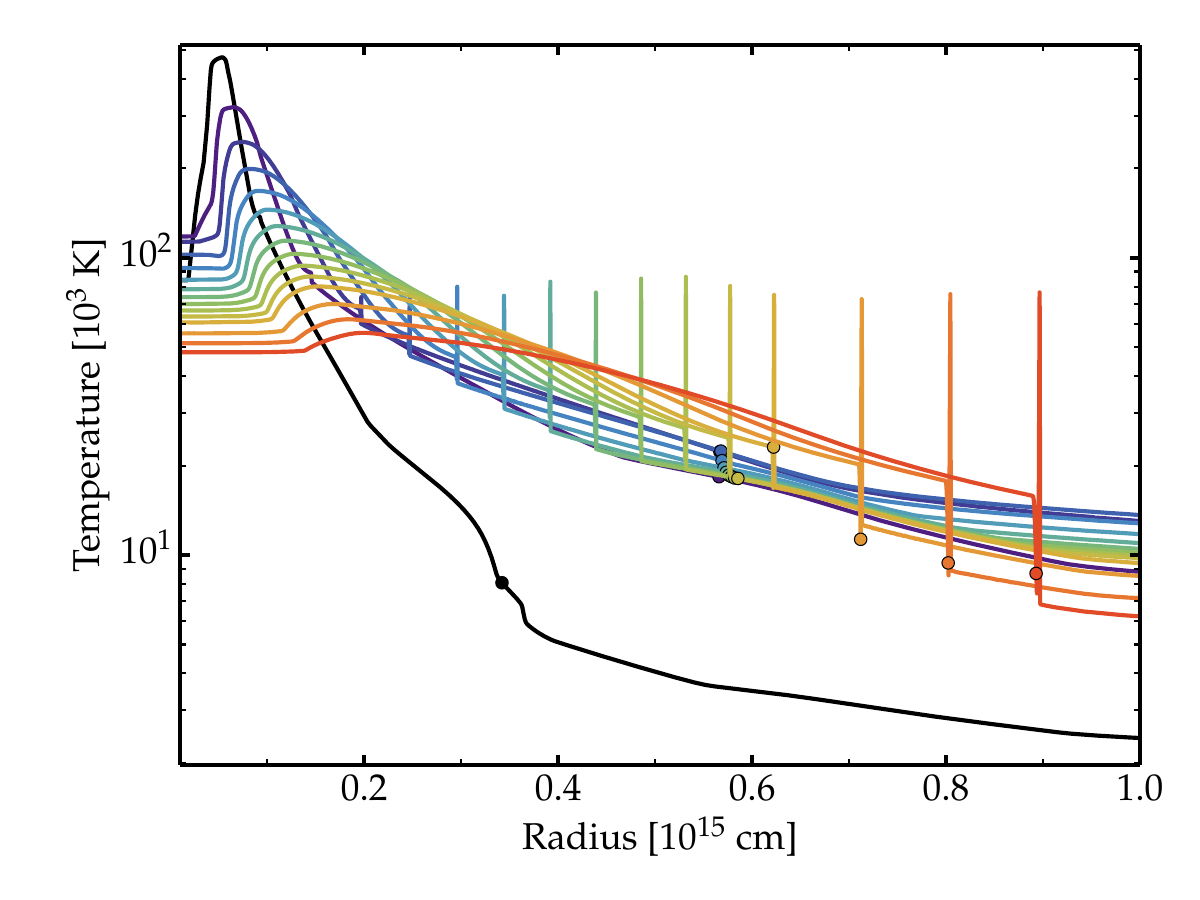}}
\subfigure[]{\includegraphics[width=0.49\textwidth]{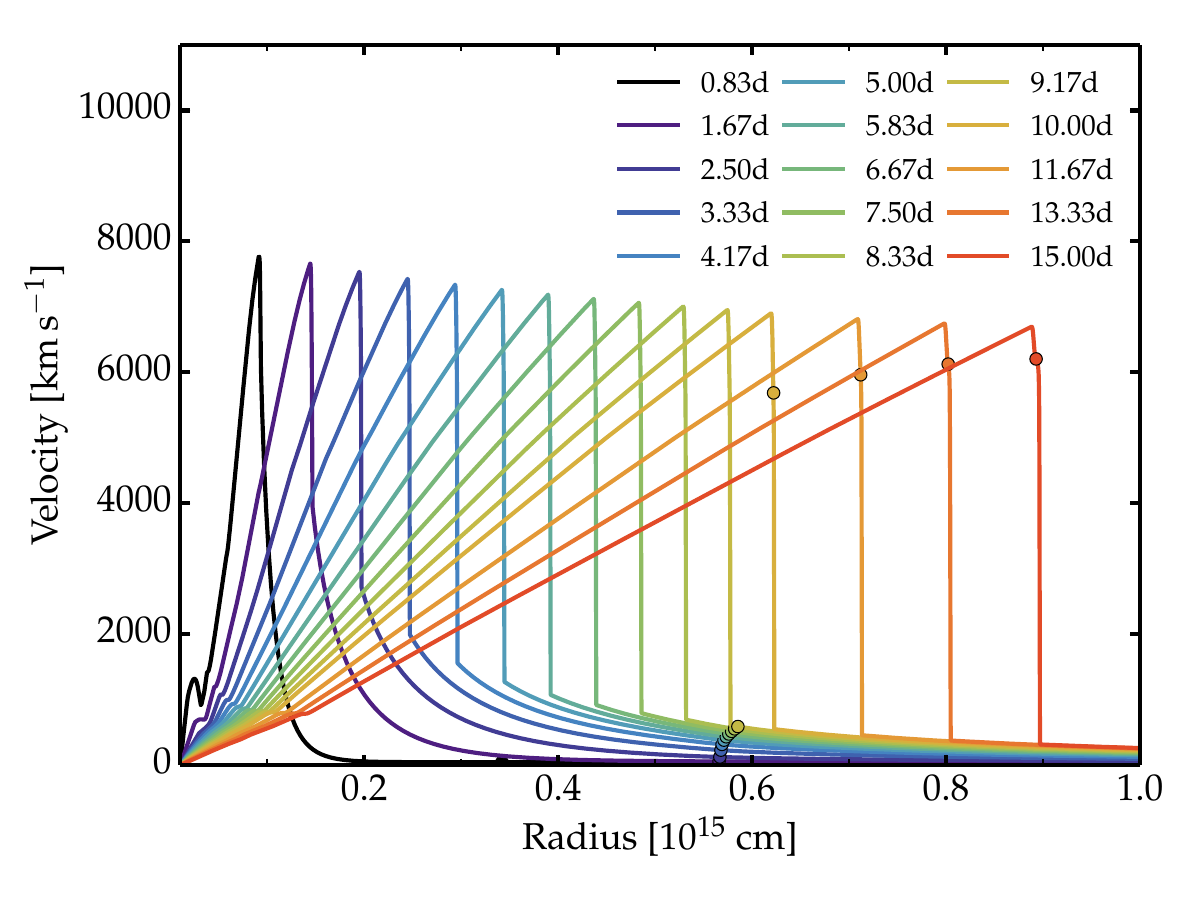}}
\caption{Time-series of the (a) luminosity, (b) density, (c) temperature, and (d) velocity versus radius for best fitting \cmfgen\ model r1w6b. Circles show the location of the photosphere, which resides in the slow moving CSM until $\sim 8$~days and after which recedes into the fast moving dense shell. The model phases begin at the onset of the radiation hydrodyanmics simulation, which is $\sim$1~hr before the shock crosses the progenitor radius (as given in the progenitor stellar model, i.e. without CSM). 
\label{fig:model_props} }
\end{figure*}

Here we present a log of optical spectroscopic observations of \sn{} in Table \ref{tab:spec_table} and a list of model properties for all \cmfgen\ simulations in Table \ref{tab:models}. Figure \ref{fig:spec_UV} presents {\it Swift} UVOT grism spectra of \sn{} and model predictions for FUV spectral features. Figure \ref{fig:model_props} shows the time-series evolution of luminosity, density, temperature and velocity as a function of radius in the r1w6b model.

\renewcommand\thetable{A\arabic{table}} 
\setcounter{table}{0}

\begin{deluxetable*}{cccccc}[h!]
\tablecaption{Optical Spectroscopy of SN~2023ixf \label{tab:spec_table}}
\tablecolumns{5}
\tablewidth{0.45\textwidth}
\tablehead{
\colhead{UT Date} & \colhead{MJD} &
\colhead{Phase\tablenotemark{a}} &
\colhead{Telescope} & \colhead{Instrument} & \colhead{Wavelength Range}\\
\colhead{} & \colhead{} & \colhead{(days)} & \colhead{} & \colhead{} & \colhead{(\AA)}
}
\startdata
2023-05-21 & 60085.20 & 2.36 & Shane & Kast & 3600--10800 \\
2023-05-21 & 60085.21 & 2.37 & Shane & Kast & 5600--7254 \\
2023-05-21 & 60085.44 & 2.61 & Shane & Kast & 3600--10800 \\
2023-05-21 & 60085.46 & 2.63 & Shane & Kast & 5600--7254 \\
2023-05-22 & 60086.20 & 3.36 & Shane & Kast & 3600--9000 \\
2023-05-22 & 60086.24 & 3.41 & APO & KOSMOS & 3600--10800 \\
2023-05-22 & 60086.31 & 3.48 & HET & LRS & 3600--7000 \\
2023-05-23 & 60087.23 & 4.39 & Shane & Kast & 3600--9000 \\
2023-05-24 & 60088.31 & 5.48 & Shane & Kast & 3600--9000 \\
2023-05-25 & 60089.20 & 6.36 & Shane & Kast & 3600--9000 \\
2023-05-27 & 60091.21 & 8.38 & Shane & Kast & 3600--9000 \\
2023-05-29 & 60093.41 & 10.58 & Shane & Kast & 3600--10800 \\
2023-05-29 & 60093.42 & 10.59 & Shane & Kast & 5600--7254 \\
2023-05-30 & 60094.25 & 11.41 & Shane & Kast & 3600--9000 \\
2023-05-31 & 60095.19 & 12.35 & Shane & Kast & 3600--9000 \\
2023-06-02 & 60097.26 & 14.43 & Shane & Kast & 3600--10800 \\
    \enddata
\tablenotetext{a}{Relative to first light (MJD 60082.83)}
\end{deluxetable*}

\begin{deluxetable*}{ccccccccc}
\tablecaption{Model Properties \label{tab:models}}
\tablecolumns{9}
\tablewidth{0pt}
\tablehead{\colhead{Name} & {$t_{\rm IIn}$} & {$\dot M$} & {$\rho_{\rm CSM,14}$$^a$} & {$R_{\rm CSM}$} & {Reference} \\
\colhead{} & \colhead{(days)} & \colhead{(M$_{\odot}$~yr$^{-1}$)} & \colhead{(g cm$^{-3}$)} & \colhead{(cm)} & \colhead{}  }
\startdata
r1w1h & $<0.3$ & 1.0e-06 & 2.7e-12 & $3\times 10^{14}$ & \cite{dessart17} \\
r1w1 & $<0.1$ & 1.0e-06 & 1.0e-16 & $1\times 10^{15}$ & \cite{dessart17} \\
r2w1 & $<0.2$ & 1.0e-06 & 9.0e-16 & $1\times 10^{14}$ & \cite{dessart17} \\
r1w4 & $1.4$ & 1.0e-03 & 1.0e-13 & $5\times 10^{14}$ & \cite{dessart17} \\
r1w5h & $0.9$ & 3.0e-03 & 5.0e-13 & $3\times 10^{14}$ & \cite{dessart17} \\
r1w5r & $1.4$ & 5.0e-03 & 5.0e-13 & $4\times 10^{14}$ & \cite{dessart17} \\
r1w6 & $3.5$ & 1.0e-02 & 1.0e-12 & $5\times 10^{14}$ & \cite{dessart17} \\
r1w6a & $5.5$ & 1.0e-02 & 1.0e-12 & $6\times 10^{14}$ & This work \\
r1w6b & $7.0$ & 1.0e-02 & 1.0e-12 & $8\times 10^{14}$ & This work \\
r1w6c & $9.0$ & 1.0e-02 & 1.0e-12 & $1\times 10^{15}$ & \cite{wjg22} \\
r1w7a & $14.0$ & 3.0e-02 & 3.0e-12 & $1\times 10^{15}$ & \cite{wjg22} \\
r1w7b & $25.0$ & 3.0e-02 & 3.0e-12 & $2\times 10^{15}$ & \cite{wjg22} \\
r1w7c & $35.0$ & 3.0e-02 & 3.0e-12 & $4\times 10^{15}$ & \cite{wjg22} \\
r1w7d & $35.0$ & 3.0e-02 & 3.0e-12 & $8\times 10^{15}$ & \cite{wjg22} \\
m1em5 & $<0.1$ & 1.0e-05 & 6.1e-16 & $1\times 10^{16}$ & \cite{Dessart23} \\
m1em4 & $<0.2$ & 1.0e-04 & 5.2e-15 & $1\times 10^{16}$ & \cite{Dessart23} \\
m1em3 & $1.0$ & 1.0e-03 & 5.4e-14 & $1\times 10^{16}$ & \cite{Dessart23} \\
m1em2 & $4.0$ & 1.0e-02 & 1.3e-12 & $1\times 10^{16}$ & \cite{Dessart23} \\
m1em1 & $15.0$ & 1.0e-01 & 1.4e-11 & $1\times 10^{16}$ & \cite{Dessart23} \\
m1em0 & $25.0$ & 1.0e+00 & 7.3e-11 & $1\times 10^{16}$ & \cite{Dessart23} \\
\enddata
\tablenotetext{a}{Density at $10^{14}$~cm}
\end{deluxetable*}

\newpage


\end{document}